\documentclass[twocolumn,tighten]{aastex6}

\usepackage{natbib}
\usepackage{epstopdf}

\newcommand{\teff}{T_\mathrm{eff}}
\newcommand{\logg}{\log\,g}
\newcommand{\feh}{\mathrm{[Fe/H]}}
\newcommand{\vt}{v_t}

\begin{document}

\shorttitle{On the Origin of $\iota$ Hor}
\shortauthors{Ram\'irez et al.}

\title{iota Horologii is unlikely to be an evaporated Hyades star}

\author{
I.\,Ram\'irez\altaffilmark{1},
D.\,Yong\altaffilmark{2},
E. Guti\'errez\altaffilmark{3},
M.\,Endl\altaffilmark{3},
D.\,L.\,Lambert\altaffilmark{3}, and
J.-D.\,Do\,Nascimento\,Jr.\altaffilmark{4,5}
}
\altaffiltext{1}{Tacoma Community College, 6501 South 19th Street, Tacoma, WA 98466 }
\altaffiltext{2}{Research School of Astronomy and Astrophysics, Australian National University, Canberra, ACT 2611, Australia}
\altaffiltext{3}{McDonald Observatory and Department of Astronomy, University of Texas at Austin, 2515 Speedway, Austin, TX 78712-1205}
\altaffiltext{4}{Departamento de Fisica, Universidade Federal do Rio Grande do Norte, CP 1641, 59072-970, Natal, Rio Grande do Norte, Brazil}
\altaffiltext{5}{Harvard-Smithsonian Center for Astrophysics, 60 Garden St., Cambridge, MA 02138}

\begin{abstract}
We present a high-precision chemical analysis of $\iota$\,Hor (iota Horologii), a planet-host field star thought to have formed in the Hyades. Elements with atomic number $6\leq Z\leq30$ have abundances that are in excellent agreement with those of the cluster within the $\pm0.01$\,dex (or $\simeq2$\,\%) precision errors. Heavier elements show a range of abundances such that about half of the $Z>30$ species analyzed are consistent with those of the Hyades, while the other half are marginally enhanced by $0.03\pm0.01$\,dex ($\simeq7\pm2$\,\%). The lithium abundance, $A$(Li), is very low compared to the well-defined $A$(Li)--$\teff$ relation of the cluster. For its $\teff$, $\iota$\,Hor's lithium content is about half the Hyades'. Attributing the enhanced lithium depletion to the planet would require a peculiar rotation rate, which we are unable to confirm. Our analysis of $\iota$\,Hor's chromospheric activity suggests $P_\mathrm{rot}=5$\,d, which is significantly shorter than previously reported. Models of Galactic orbits place $\iota$\,Hor hundreds of parsecs away from the cluster at formation. Thus, we find the claim of a shared birthplace very difficult to justify.
\end{abstract}

\keywords{stars: abundances --- stars: fundamental parameters 
 --- stars: individual ($\iota$ Hor) --- stars: planetary systems --- open clusters and associations: individual (Hyades)}

\section{Introduction}

Most open clusters are expected to dissolve into the Milky Way's disk on time scales of a few hundred million years \citep{janes88}. The Hyades, while easily identified as an open cluster in the night sky, have an age of about 650\,Myr \cite[e.g.,][]{perryman98,degennaro09}. This implies that, most likely, this cluster is already ``evaporating;'' some of its stars might now be in the field, far from their parent cluster. Indeed, a fraction of the stars in the so-called Hyades stream might be former Hyades cluster stars \citep{pompeia11}, even though it is more likely that the majority of them have been led to have Hyades-like Galactic space velocities by dynamical processes \citep{famaey07}.

One candidate star of the Hyades stream is of particular interest: $\iota$\,Hor (HR\,810, HD\,17051, HIP\,12653), a naked-eye ($V=5.4$\,mag) young solar-type star visible in the Southern Hemisphere sky, which is known to host a planet. Analysis of high-precision radial velocity data by \cite{kurster00} first suggested the presence of a $\simeq2\,M_\mathrm{Jup}$ minimum-mass planet at $\simeq1$\,AU from the star \cite[see also][]{zechmeister13}. At the time of its discovery, this planet had the most Earth-like orbit known. Exoplanets in the Hyades appeared elusive \citep{paulson04} until recent discoveries \citep{quinn14,mann16}. If a true Hyades member, $\iota$\,Hor would add significantly to the planet statistics of this cluster. Note, however, that planet-host stars in open clusters are not particularly rare \cite[e.g.,][]{quinn12,brucalassi14,malavolta16}.

\cite{montes01} studied the possibility that $\iota$\,Hor belongs to the Hyades supercluster (or stream). They provide Hyades supercluster mean $U,V,W$ velocities (heliocentric) of $-39.7$, $-17.7$, $-2.4$ km/s, respectively, and used the following values for $\iota$\,Hor: $U=-31.27\pm0.31$\,km/s, $V=-16.44\pm0.69$\,km/s, $W=-7.67\pm1.11$\,km/s. \citeauthor{montes01} applied two of the criteria by \cite{eggen95} to determine whether a star is moving towards the convergent point of its pressumed supercluster and found that $\iota$\,Hor satisfies one of them.

Using HARPS spectra, \cite{vauclair08} performed a study of acoustic oscillations in $\iota$\,Hor, which confirmed the stars' young age and super-solar metallicity, both in good agreement with those of Hyades stars. In addition, they measured a helium abundance for $\iota$\,Hor and claimed a close match to that of the Hyades, as measured by \cite{lebreton01} using model fits to the mass-luminosity relation of five binary systems in the cluster. These results led \citeauthor{vauclair08} to the suggestion that $\iota$\,Hor was formed in the Hyades, even though today it is about 40\,pc away from it and the tidal radius of the Hyades is only $\simeq10$\,pc \citep{perryman98}.

Our goal in this paper is to re-visit \citeauthor{vauclair08}'s claim of a $\iota$\,Hor--Hyades connection using more recent and therefore more precise stellar parameter and chemical abundance analysis, in addition to other relevant complementary information on chromospheric activity, rotation, and helium abundance. Instead of looking at only one element besides helium, namely iron, we perform a high-precision, multi-element investigation, including the very important element lithium.

\section{The Challenge of Chemical Tagging}

If formed from the same gas cloud, stars should have identical chemical composition, as long as the cloud was well mixed when stars formed. Thus, to confirm a field star as a former cluster member, one could in principle look into its detailed composition and determine whether it matches that of the cluster. In a way, this is the fundamental premise of ``chemical tagging'' \cite[e.g.,][]{freeman02,hogg16}.

High-resolution spectroscopy allows us to investigate elemental abundances in great detail. However, systematic uncertainties in standard model atmosphere analyses \cite[see, e.g., the review by][]{asplund05:review} limit our ability to measure chemical abundances at an appropriate level for chemical tagging to be a reliable tool. Of particular interest to our study is the fact that in some of the most comprehensive multi-element abundance analyses of nearby disk stars, the Hyades do not stand out in a major way \cite[e.g.,][]{reddy06}.

By performing a strict differential analysis of stars that are nearly identical to each other (``stellar twins''), systematic uncertainties can be essentially removed, and thus the use of very high-quality spectra guarantees extremely precise measurements of stellar parameters and relative abundances.\footnote{In practice, this is attainable when the effective temperatures, surface gravities, and iron abundances of the stars are approximately within 100\,K, 0.1, and 0.1\,dex of each other, respectively.} Indeed, analysis of twin-star spectra with resolution $R\gtrsim40\,000$ and signal-to-noise ratio $S/N\gtrsim300$ now allows us to determine chemical abundances with 0.01\,dex precision.\footnote{In quantities such as [X/H], where X is a chemical element with number density $n_\mathrm{X}$ and $\mathrm{[X/H]}=A(\mathrm{X})-A(\mathrm{X})^\odot$, with $A(\mathrm{X})=\log(n_\mathrm{X}/n_\mathrm{H}) + 12$.} Thus, these new efforts in high-precision spectroscopy \citep[e.g.,][]{melendez09:twins,ramirez14:bst,ramirez14:harps,ramirez15,spina16} can have an important and significant impact on the prospects for chemical tagging. The reality of the very small uncertainties achieved in this type of analyses has been investigated by \cite{bedell14} and supported by comparison with other fully independent high-precision studies such as those by \cite{nissen15}.

No comprehensive study of a large sample (several hundred to thousands) of nearby stars at 0.01\,dex precision exists yet, but today we can realistically examine small samples of interest like a handful of Hyades stars and $\iota$\,Hor as a proof-of-concept test. Note, however, that being a bright star, $\iota$\,Hor has been included in a number of large-sample, multi-element abundance works before \cite[e.g.,][]{adibekyan12,gonzalez-hernandez13,bensby14,ramirez14:bst}.

\citet[][hereafter L16]{liu16} have performed a high-precision spectroscopic study of a sample of Hyades stars, and it is thus a natural starting point for the test discussed above. L16 found that the Hyades open cluster is inhomogeneous. Its stars have a range of abundances that extend over 0.1\,dex for most species they analyzed. L16 insisted that this inhomogeneity in chemical composition cannot be explained by residual errors in the analysis. Correlations between enhancements and depletions were observed for most elements analyzed and they were shown to be independent of stellar parameters, minimizing potential systematic errors.

Here we employ high quality spectra of $\iota$\,Hor and one of its twin stars in the Hyades (HD\,28635) to explore the possibility that it was formed in said cluster. By measuring the composition of $\iota$\,Hor relative to that of its Hyades twin, we are able to compare it to other Hyades cluster stars at very high precision as well because the star chosen as comparison is included in the L16 study.

\section{Spectroscopic Data}\label{s:data}

\begin{table}
\caption{ESO/HARPS Archive Spectra Used in the Chemical Abundance Analysis}
\label{t:archive}
\centering
\begin{tabular}{lcrcr}\hline\hline
Star & Run ID & $n$\tablenotemark{$a$} & $t$\tablenotemark{$b$} & $S/N$\tablenotemark{c} \\ \hline
{\bf $\iota$\,Hor} &               &  {\bf 66} &  {\bf 3.97} & {\bf 1217} \\
             & 091.C-0853(A) &  25 &  2.83 &      \\
             & 60.A-9036(A)  &   8 &  0.22 &      \\
             & 60.A-9700(G)  &  33 &  0.92 &      \\
{\bf HD\,28635}      &               &  {\bf 35} &  {\bf 1.33} &  {\bf 283} \\
             & 075.D-0614(A) &  34 &  1.13 &      \\
             & 094.D-0596(A) &   1 &  0.19 &      \\
\hline
\end{tabular}
\begin{flushleft}
\tablenotetext{a}{Number of spectra}
\tablenotetext{b}{Total exposure time in hours}
\tablenotetext{c}{Signal-to-noise ratio {\it per pixel} measured near 6200\,\AA.}
\end{flushleft}
\end{table}

Most of the analysis presented in this work is based on high-resolution HARPS spectra ($R=120\,000$) obtained from the ESO archive (see Table~\ref{t:archive}). 
Although a number of Hyades stars have been observed with HARPS (in particular within Run ID 094.D-0596(A), PI.\ L.~Pasquini), only a fraction are twins of $\iota$\,Hor and few have spectra of sufficient quality. The best star that we could find in the ESO/HARPS archive for our purposes is HD\,28635, which is one of the targets of the L16 study. 

In addition to the HARPS data, we used the spectra employed in L16 to measure the lithium abundances of that sample of Hyades stars and compare them to that of $\iota$\,Hor. 
The L16 data were taken with the Tull coud\'e spectrograph on the 2.7\,m Telescope at McDonald Observatory and they have $R=60\,000$, with a typical $S/N\simeq350-400$ per pixel. 
We also used a UVES spectrum of $\iota$\,Hor found in the ESO archive (Program ID 084.D-0965(A)) to measure its oxygen abundance from the triplet lines at 777\,nm, a wavelength region not available in the HARPS spectrum. 

\section{Elemental Abundances} \label{s:abundances}

Stellar parameters and elemental abundances of $\iota$\,Hor were measured differentially with respect to its twin star HD\,28635 using standard 1D-LTE model atmosphere analysis. Since HD\,28635 is included in the L16 study, we replicated their analysis so that $\iota$\,Hor could be simply added to that sample for further investigation. Thus, we employed the same model atmospheres (Kurucz' ``odfnew'' grid; \citealt{castelli03}) and differential excitation/ionization balance of iron lines to determine the atmospheric parameters. Abundances were calculated using equivalent widths and curve-of-growth analysis with the ``abfind'' driver of MOOG \citep{sneden73}.\footnote{\url{http://www.as.utexas.edu/~chris/moog.html}} We employed the $q^2$ Python code \citep{ramirez14:harps}\footnote{\url{https://github.com/astroChasqui/q2}} for the manipulation of MOOG input and output files instead of the equivalent IDL routines used in L16. A multitude of tests carried out before the publication of L16 by I.\ Ram\'irez and F.\ Liu ensured that these tools provide identical results given equal input data. Thus, the results presented in this paper for $\iota$\,Hor are on the same scale as those given in the L16 paper.

The differential stellar parameters we obtained for $\iota$\,Hor are given in Table~\ref{t:parameters}. The parameters of the reference star, HD\,28635, are also listed there, but without error, since they were assumed fixed for these calculations. The parameters adopted for the comparison star are identical to those derived by L16.

\begin{table}
\caption{Stellar Parameters}
\label{t:parameters}
\centering
\scriptsize
\begin{tabular}{lcccc}\hline\hline
Star & $\teff$ & $\logg$ & $\feh$ & $\vt$ \\
     & (K) & [cgs] & & (km/s) \\ \hline
HD\,28635 & 6278 & 4.53 & 0.156 & 1.34 \\
$\iota$\,Hor & $6232\pm8$ & $4.55\pm0.02$ & $0.140\pm0.005$ & $1.34\pm0.01$ \\ \hline
\end{tabular}
\end{table}


The linelists employed in L16 and in this work are almost identical. The small differences stem primarily from the dissimilar wavelength coverage available in each set of data. As in L16, we measured the abundances of C, O, Na, Mg, Al, Si, S, Ca, Sc, Ti, V, Cr, Mn, Fe, Co, Ni, Cu, Zn, and Ba. In addition, we determined the abundances of Sr, Y, Zr, La, Ce, Nd, Sm, and Eu. The linelists used for these species, and hyperfine structure parameters, when necessary, were adopted from \cite{reddy06} and \cite{melendez14:18sco}.

\begin{table}
\caption{Elemental Abundances\tablenotemark{$a$}}
\label{t:abundances}
\centering
\tiny
\begin{tabular}{rlRrr}\hline\hline
\tabletypesize{\tiny}
$Z$\tablenotemark{$b$} & Species & $\Delta\mathrm{[X/H]}$ & $\sigma(\Delta\mathrm{[X/H]})$ & $n$\tablenotemark{$c$} \\ \hline
  6.0 & CI   & -0.021 & 0.009 & 3 \\ 
106.0 & CH   & -0.020 & 0.008 & 1 \\ 
  8.0 & OI   & -0.016 & 0.014 & 3 \\ 
 11.0 & NaI  & -0.013 & 0.011 & 4 \\ 
 12.0 & MgI  & -0.012 & 0.008 & 2 \\ 
 13.0 & AlI  & -0.008 & 0.006 & 3 \\ 
 14.0 & SiI  &  0.000 & 0.004 & 22 \\ 
 16.0 & SI   & -0.017 & 0.013 & 4 \\ 
 20.0 & CaI  & -0.003 & 0.006 & 16 \\ 
 21.0 & ScI  & -0.017 & 0.006 & 4 \\ 
 21.1 & ScII & -0.004 & 0.008 & 7 \\ 
 22.0 & TiI  & -0.014 & 0.008 & 48 \\ 
 22.1 & TiII & -0.003 & 0.008 & 17 \\ 
 23.0 & VI   & -0.017 & 0.008 & 16 \\ 
 24.0 & CrI  & -0.020 & 0.006 & 33 \\ 
 24.1 & CrII & -0.008 & 0.008 & 8 \\ 
 25.0 & MnI  & -0.002 & 0.007 & 11 \\ 
 26.0 & FeI  & -0.012 & 0.006 & 87 \\ 
 26.1 & FeII & -0.017 & 0.009 & 17 \\ 
 27.0 & CoI  & -0.022 & 0.006 & 15 \\ 
 28.0 & NiI  & -0.013 & 0.005 & 55 \\ 
 29.0 & CuI  & -0.001 & 0.008 & 2 \\ 
 30.0 & ZnI  & -0.005 & 0.008 & 2 \\ 
 38.0 & SrI  & -0.001 & 0.008 & 1 \\ 
 39.1 & YII  &  0.020 & 0.010 & 6 \\ 
 40.1 & ZrII & -0.007 & 0.008 & 1 \\ 
 56.1 & BaII &  0.022 & 0.008 & 3 \\ 
 57.1 & LaII & -0.006 & 0.008 & 2 \\ 
 58.1 & CeII &  0.028 & 0.008 & 6 \\ 
 60.1 & NdII & -0.006 & 0.009 & 3 \\ 
 62.1 & SmII &  0.012 & 0.008 & 4 \\ 
 63.1 & EuII &  0.013 & 0.008 & 2 \\ 
\hline
\end{tabular}
\begin{flushleft}
\tablenotetext{a}{$\iota$\,Hor -- HD\,28635.}
\tablenotetext{b}{Species code as defined in MOOG.}
\tablenotetext{c}{Number of spectral lines employed.}
\end{flushleft}
\end{table}

Our derived differential elemental abundances and their errors are given in Table~\ref{t:abundances}. As in previous high-precision abundance work, the errors provided there correspond to the line-to-line standard error added in quadrature with the uncertainties obtained by propagating the errors in stellar parameters. For species with fewer than three spectral lines available, the error was assumed to be 0.008\,dex, which is the average uncertainty of species for which 3 or more lines are available. 

For comparison with the Hyades sample from L16, the abundances we measured for $\iota$\,Hor relative to HD\,28635 were corrected for the relative abundances between this twin star and the reference star in L16, HD\,25825, using the values derived by L16. Of the $Z>30$ elements analyzed in this paper, only barium is included in L16. Spectral lines of these heavy metals are often weak and blended. Therefore they require extremely high quality spectra to be properly measured. The HARPS data used in this work for the $\iota$\,Hor -- HD\,28635 comparison have a higher quality than the McDonald spectra used in L16, primarily because of the significantly higher spectral resolution of the former. 

\begin{table}
\caption{Sr, Y, and Ce abundances of L16's Hyades stars\tablenotemark{$a$}}
\label{t:sryce_liu16}
\centering\scriptsize
\begin{tabular}{lRcRcRc}\hline\hline
Star & $\Delta\mathrm{[Sr/H]}$ & err & $\Delta\mathrm{[Y/H]}$ & err & $\Delta\mathrm{[Ce/H]}$ & err \\ \hline
HD25825 & 0.000  &       & 0.000  &       & 0.000  &       \\
HD26736 & 0.069  & 0.027 & 0.045  & 0.024 & 0.049  & 0.031 \\
HD26756 & 0.049  & 0.033 & 0.037  & 0.039 & 0.046  & 0.017 \\
HD26767 & 0.061  & 0.017 & 0.064  & 0.011 & 0.038  & 0.040 \\
HD27282 & 0.085  & 0.039 & 0.058  & 0.036 & 0.056  & 0.021 \\
HD27406 & 0.024  & 0.029 & 0.030  & 0.019 & 0.009  & 0.019 \\
HD27835 & 0.038  & 0.023 & 0.033  & 0.020 & 0.026  & 0.015 \\
HD27859 & -0.038 & 0.021 & -0.023 & 0.016 & -0.039 & 0.014 \\
HD28099 & 0.069  & 0.033 & 0.044  & 0.022 & 0.000  & 0.021 \\
HD28205 & 0.058  & 0.026 & 0.060  & 0.020 & 0.032  & 0.020 \\
HD28237 & -0.014 & 0.028 & -0.009 & 0.024 & -0.014 & 0.020 \\
HD28344 & 0.047  & 0.016 & 0.056  & 0.017 & 0.057  & 0.012 \\
HD28635 & 0.021  & 0.024 & 0.042  & 0.020 & -0.003 & 0.021 \\
HD28992 & 0.011  & 0.021 & 0.016  & 0.017 & -0.017 & 0.016 \\
HD29419 & 0.031  & 0.023 & 0.030  & 0.026 & 0.006  & 0.018 \\
HD30589 & 0.064  & 0.022 & 0.055  & 0.031 & 0.034  & 0.018 \\
\hline
\end{tabular}
\begin{flushleft}
\tablenotetext{a}{Star -- HD\,25825.}
\end{flushleft}
\end{table}

We attempted to measure the abundances of all $Z>30$ elements listed in Table~\ref{t:abundances} for the Hyades stars of the L16 work, but we were only able to measure the abundances of Sr, Y, and Ce using equivalent widths. Moreover, not all the lines used to measure the $\iota$\,Hor minus HD\,28635 relative Y and Ce abundances had sufficient quality in the McDonald spectra for us to trust the measured $EW$ values. Thus, these abundances are based on a lower number of spectral lines than used for $\iota$\,Hor (1 Sr line, 3 Y lines, and 2 Ce lines). The abundances of Sr, Y, and Ce we measured in the McDonald spectra for the Hyades stars from L16 are given in Table~\ref{t:sryce_liu16}. Note that these abundances are given relative to the reference star in the L16 study, which is HD\,25825, not the $\iota$\,Hor twin HD\,28635, but corrections were applied whenever necessary in the analysis. The stellar parameters $\teff$, $\logg$, $\feh$, and $\vt$ used in these calculations are those derived by L16, which are somewhat less precise that our $\iota$\,Hor -- HD\,28635 relative parameters. Considering all these factors, it is not surprising that our errors for the Sr, Y, and Ce abundances in Table~\ref{t:sryce_liu16} are larger than the errors listed in Table~\ref{t:abundances}.

\begin{table}
\caption{Lithium Abundances}
\label{t:li}
\centering
\begin{tabular}{lcrcc}\hline\hline
Star & $\teff$ & $EW$ & $A$(Li) & $P_\mathrm{rot}$ \\
 & (K) & (m\AA) & & (days) \\ \hline
$\iota$\,Hor & 6232 & 39.4 & 2.52 & $\simeq5$ \\
HD25825  & 6094 &  78.1 & 2.78 & 7.41 \\ 
HD26736  & 5896 &  65.3 & 2.55 & 8.34 \\ 
HD26756  & 5760 &  38.4 & 2.17 & \nodata \\ 
HD26767  & 5944 &  79.6 & 2.69 & 6.10 \\ 
HD27282  & 5654 &  26.0 & 1.90 & 8.91 \\ 
HD27406  & 6225 &  91.7 & 2.96 & 5.47 \\ 
HD27835  & 6070 &  62.9 & 2.65 & \nodata \\ 
HD27859  & 6034 &  76.7 & 2.72 & 7.81 \\ 
HD28099  & 5819 &  58.7 & 2.43 & 8.66 \\ 
HD28205  & 6306 & 103.7 & 3.10 & 5.87 \\ 
HD28237  & 6238 &  77.7 & 2.89 & 5.13 \\ 
HD28344  & 6074 &  90.5 & 2.85 & 7.41 \\ 
HD28635  & 6278 &  95.6 & 3.02 & \nodata \\ 
HD28992  & 5968 &  80.6 & 2.70 & 8.72 \\ 
HD29419  & 6180 &  87.3 & 2.91 & \nodata \\ 
HD30589  & 6142 &  96.8 & 2.94 & \nodata \\ 
\hline
\end{tabular}
\end{table}

The lithium abundance of $\iota$\,Hor and all stars in the L16 sample, including HD\,28635, were measured using the equivalent widths ($EW$s) of the 6707.8\,\AA\ lithium doublet, available in our Tull/2.7\,m McDonald spectra, and the lithium abundance interpolator provided by the INSPECT project.\footnote{\url{http://www.inspect-stars.com}} The latter uses a grid of non-LTE lithium abundances abundances calculated by \cite{lind09}. 
Our 6707.8\,\AA\ Li doublet line $EW$ measurements and the lithium abundances inferred from them are given in Table~\ref{t:li}. We also include in this table effective temperatures measured by L16. While the $\teff$ values themselves might have oridinary accuracy, the differences in $\teff$ are extremely precise, having errors of order 10\,K. Propagating errors in stellar parameters and taking into account the noise in the spectra we estimate that our lithium abundances have an average error of 0.05\,dex.

\begin{figure*}
\centering
\includegraphics[width=17.5cm, trim={0.5cm 0.5cm 0 0}]{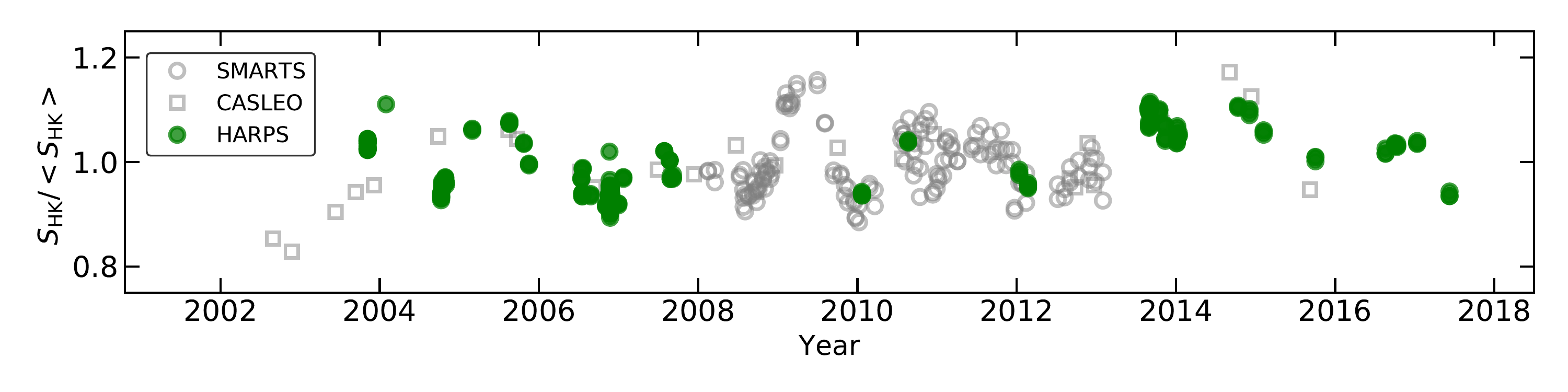}
\caption{Long-term variability of the Ca\,\textsc{ii} H\&K chromospheric activity index normalized to its average. Measurements made on three different data sets are shown: SMARTS (open circles), CASLEO (squares), and HARPS (filled circles).}
\label{f:shk}
\end{figure*}

\section{Rotation and Activity} \label{s:rotation}

Table~\ref{t:li} includes the rotation periods ($P_\mathrm{rot}$) measured by A.\,Kundert and published in her thesis.\footnote{\url{https://discoverarchive.vanderbilt.edu/bitstream/handle/1803/5108/Kundert-thesis.pdf}} These rotation periods were determined using the rotational modulation of sunspots on light curves observed by the All Sky Automated Survey.\footnote{\url{http://www.astrouw.edu.pl/asas}} The $P_\mathrm{rot}$ measurements by Kundert are in excellent agreement with previously published values, where available. In fact, based on a comparison with literature values, these rotation periods are estimated to have a 1\,$\sigma$ error of about 0.5\,days.

The rotation period of $\iota$\,Hor was reported to be 7.9 days by \cite{saar97:south}, who employed chromospheric activity measurements, while \cite{boisse11} report periods in the range from 7.9 to 8.4 days based on a statistical analysis of radial velocity jitter. In a different activity study, \cite{saar97:helium} find a rotation period of 8.6 days, while \cite{metcalfe10} obtained a value of $8.5\pm0.1$ days in their own chromospheric activity periodogram.

The range of previously published rotation periods for $\iota$\,Hor goes from 7.9 to 8.6 days. A periodogram of radial velocity residuals by \cite{zechmeister13}, after subtracting the planet signal, reveals peaks at 5.7 days and $\simeq8$ days, but these authors do not comment on the significance of the shorter period. Ignoring for a moment this lower value, previously published values for the $P_\mathrm{rot}$ of $\iota$\,Hor have an average of $8.2\pm0.3$\,days.

Detailed analysis of stellar activity indices such as chromospheric Ca\,\textsc{ii} H\&K and coronal X-ray emission lines suggest a magnetic cycle of about 1.6\,yr for $\iota$\,Hor \citep{metcalfe10,sanz-forcada13}. More recently, \cite{flores17} have found evidence for a long-term cycle of about 5 years,  emphasizing that both the short-term and long-term cycles are irregular, just like the Sun's ``11-year'' cycle, which lasts anywhere between about 9 and 14 years \citep{hathaway15}.

Using all available HARPS spectra in the ESO archive as of June 2017, we computed relative $S_\mathrm{HK}$ indices in the standard manner. $H$ and $K$ fluxes were first determined by integrating the count values around the H and K lines multiplied by triangular filters 2.1\,\AA-wide centered around the lines' cores (3968.47 and 3933.66\,\AA, respectively). $R$ and $V$ continuum fluxes were then calculated by integration of counts in the $4001.07\pm20$\,\AA\ and $3901.07\pm20$\,\AA\ windows, respectively. Finally, we computed an uncalibrated chromospheric activity index as $S_\mathrm{HK}=(H+K)/(R+V)$.

Figure~\ref{f:shk} shows the variability of the chromospheric activity indices we calculated relative to their average value as a function of time. We include the measurements by \citet[][SMARTS]{sanz-forcada13} and \citet[][CASLEO]{flores17} in this plot as well, after scaling them to our values using measurement sets taken within 10 days of each other (i.e., by ensuring that the overlaps give the same average $S_\mathrm{HK}$).

\begin{figure}
\includegraphics[width=8.6cm, trim={2.0cm 0.5cm 0 1.0cm}]{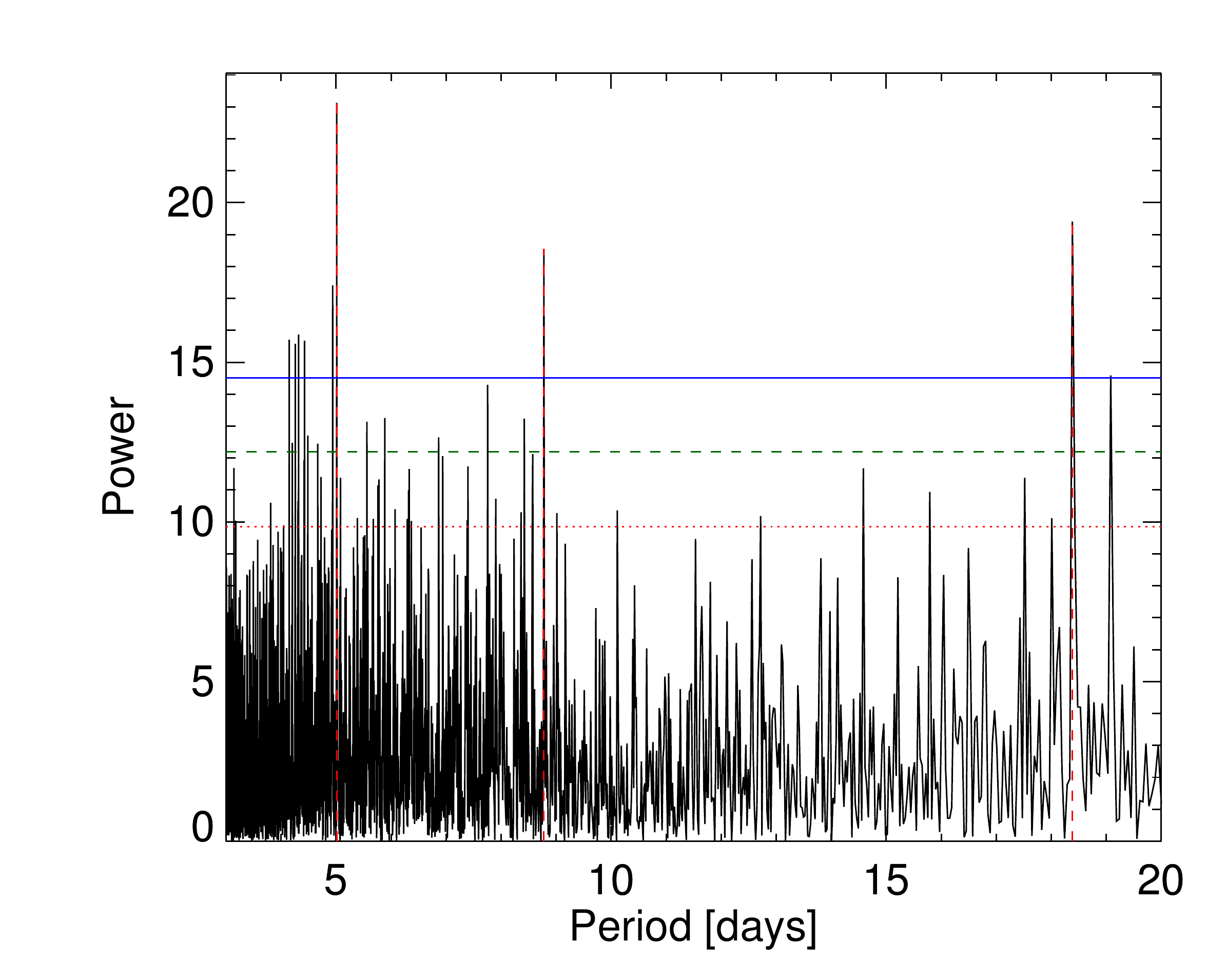}
\caption{Lomb Scargle periodogram of $\iota$\,Hor's chromospheric activity data. The most prominent peak is at a period of 5.0 days and it has a false alarm probability (FAP) much lower than 0.1\,\%. The three horizontal lines indicate FAPs of 0.1\,\% (solid line), 1\,\% (dashed line) and 10\,\% (dotted line).}
\label{f:ls}
\end{figure}

\begin{figure}
\centering
\includegraphics[width=8.2cm, trim={0.5cm 0.5cm 0 0}]{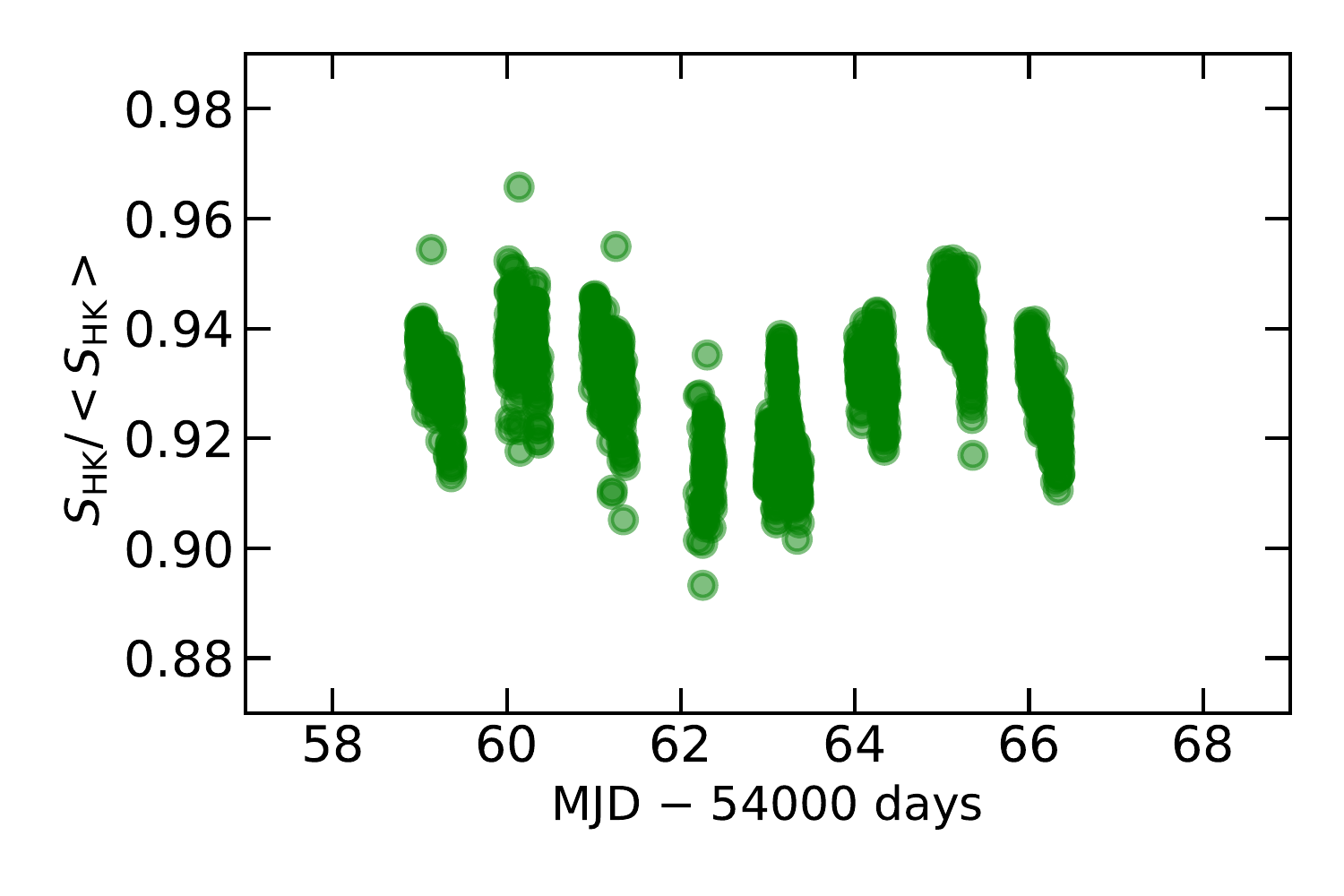}
\caption{Short-term variability of $\iota$\,Hor's Ca\,\textsc{ii} H\&K emission index at a modified Julian date (MJD) of about 54058 days.}
\label{f:shk_zoom}
\end{figure}

A generalized Lomb-Scargle (LS) periodogram analysis \citep{lomb76,scargle82,horne86} of our complete $S_\mathrm{HK}$ dataset, averaged by day to prevent generating a forest of peaks at short frequency, shows maximum power in a 5.0-day peak (Figure~\ref{f:ls}). The second highest peak is at 18.4\,d, while the third highest peak is at 8.8\,d. The 5-day periodicity appears obvious in a subset of data densely concentrated around $\mathrm{MJD}=54060$\,days (Figure~\ref{f:shk_zoom}). Since this analysis implies a $P_\mathrm{rot}=5.0$\,d that is significantly shorter than virtually all previously reported values, it is necessary to investigate its reliability.

First, we were able to reproduce the long-term activity periods quoted in the literature: 1.6 and 4.6 years. Then, we calculated LS periodograms in 100-day windows, determined the maximum in each window, and plotted them against MJD (sliding LS analysis, as in \citealt{clarkson03}), finding consistently a short period for the data, particularly during minima of chromospheric activity. We were also able to replicate \cite{metcalfe10} analysis of SMARTS data. We produced a LS periodogram identical to that shown in their Figure~2, which shows only periods longer than about 5.2\,days. Interestingly, when we extend the range of this periodogram to include shorter periods, we discover a very narrow peak at 5\,days. This peak has slightly more power than the 8.5-day peak that \cite{metcalfe10} reported as the star's rotation period. Finally, we calculated a LS periodogram excluding the data points shown in Figure~\ref{f:shk_zoom}, which reveals clearly the 5-day periodicity, and found that the 5-day period peak does not lose any significant power in the periodogram. In other words, the power in that 5-day period is not entirely dependent on the subset of data shown in Figure~\ref{f:shk_zoom}

The tests discussed above strongly support our finding of a 5-day periodicity in the chromospheric emission data of $\iota$\,Hor. We find no real reason to exclude this 5-day period as a posible value for the rotation rate of the star. In fact, it appears that this shorter period is more reliable than the $\simeq8$-day period previously reported by several authors. Thus, hereafter we adopt $P_\mathrm{rot}=5$\,days for $\iota$\,Hor.

\begin{figure*}
\includegraphics[width=17.5cm, trim={0 0.5cm 0 0}]{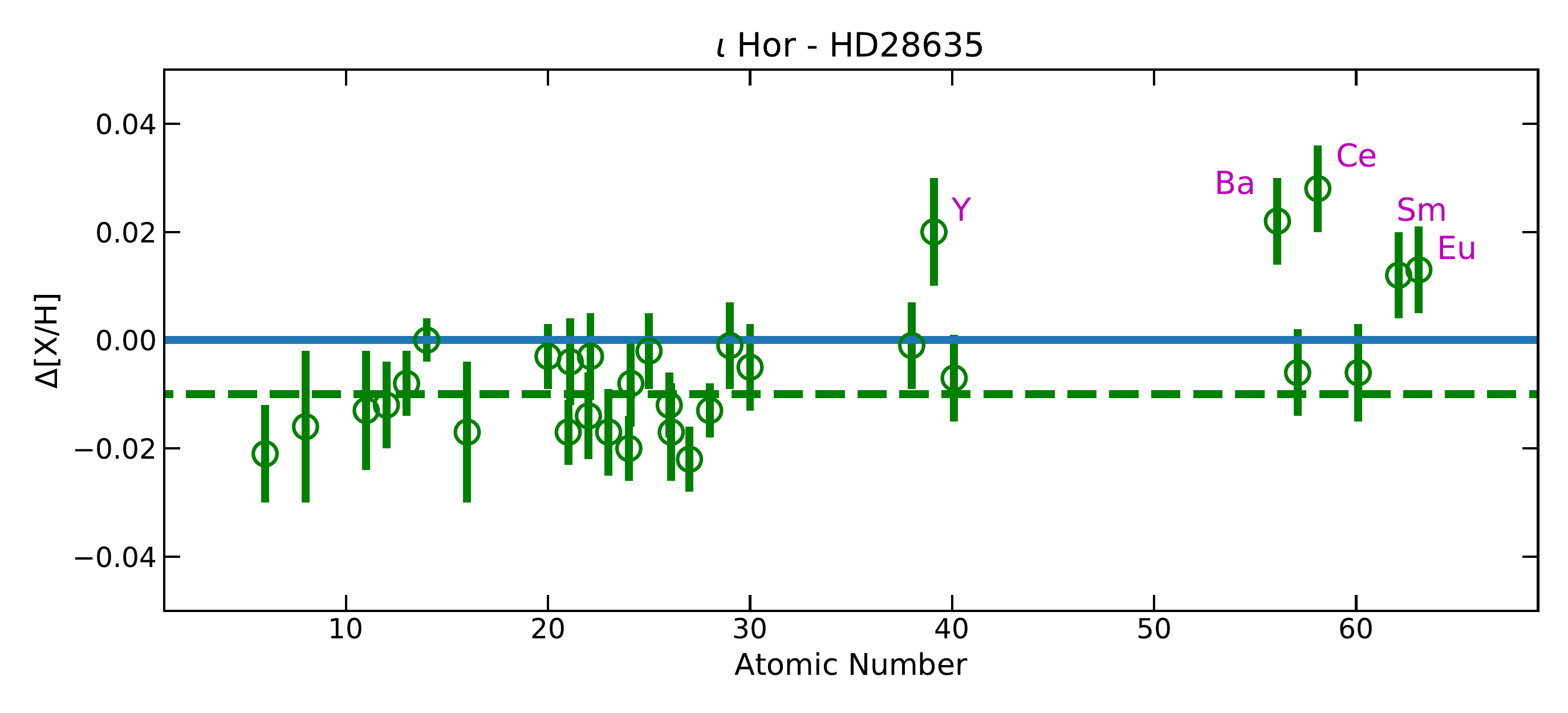}
\caption{Elemental abundance difference as a function of atomic number in the $\iota$\,Hor minus HD\,28635 sense.}
\label{f:delta_xh}
\end{figure*}

\section{Discussion}

\subsection{Light and Heavy Elements}

The elemental abundances of $\iota$\,Hor, measured with respect to its Hyades twin star HD\,28635 are shown in Figure~\ref{f:delta_xh}. For the majority of species analyzed, there is a clear offset of about $-0.01$\,dex with respect to HD\,28635. In fact, the weighted average $\Delta$[X/H] value for $Z\leq30$ elements is $-0.010\pm0.007$\,dex. The element-to-element scatter in this case could be fully explained by the measurement errors, which are about 0.008\,dex.

The situation for $Z>30$ elements is more complex. On average, we find $\Delta$[X/H]$=+0.008\pm0.013$\,dex, which shows an element-to-element scatter larger than the errors. About half of the $Z>30$ elements have abundances that are consistent with the $-0.010$\,dex offset of the low-$Z$ elements, but the other half show abundances enhanced by about $0.03\pm0.01$\,dex, which corresponds to about 7\,\%. The latter consists of elements Y, Ba, Ce, Sm, and Eu.

\begin{figure}
\includegraphics[width=8.0cm, trim={0 0.2cm 0 0}]{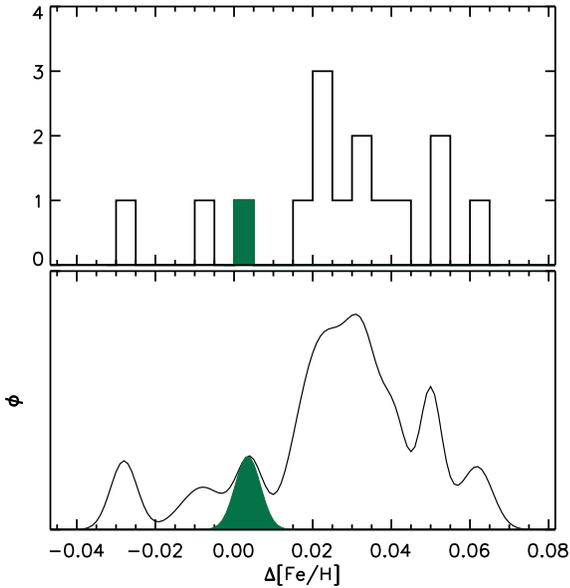}
\caption{Regular (upper panel) and generalized (lower panel) histograms of iron abundances for Hyades stars included in L16. The filled histograms represent $\iota$\,Hor.}
\label{f:feh_hist}
\end{figure}

L16 found that the Hyades open cluster is inhomogeneous; Hyades stars have an ``allowed'' range of elemental abundances. Indeed, Figure~\ref{f:feh_hist} shows that the $\Delta$[Fe/H] values measured by L16 relative to their reference star HD\,25825 have a range of $\sim0.1$\,dex. The abundances we measured for $\iota$\,Hor, after correcting for the different reference stars employed in this work and in L16, are on the low abundance side of this distribution, but lie well within the allowed range of the Hyades. The same is true for all elemental abundances measured in this work and in L16, with only one possible exception, namely barium (see below).

\begin{figure}
\includegraphics[width=8.6cm, trim={0.1cm 0.1cm 0 0}]{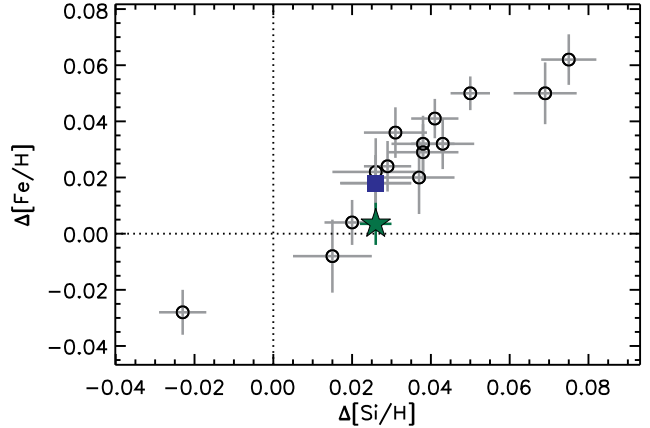}
\caption{Silicon versus iron abundances of Hyades stars (open circles) relative to HD\,25826, the reference star in the L16 study. The location of $\iota$\,Hor in this plot is indicated with the star symbol while that of its Hyades twin, HD\,28635, is shown with the filled square.}
\label{f:si_fe}
\end{figure}

One could therefore argue that the elemental abundances of $\iota$\,Hor are consistent with those of the Hyades given their intrinsic scatter. Note, however, that L16 also showed that the relative abundances of different elements in Hyades stars are correlated as opposed to randomly distributed. For example, in Figure~\ref{f:si_fe} we show the correlation between the relative abundances of iron and silicon in the Hyades sample of L16. 
Although $\iota$\,Hor is on the low metallicity side of the Hyades elemental abundance distribution, it is in excellent agreement with the overall Hyades trend.

\begin{figure}
\includegraphics[width=8.6cm, trim={0.1cm 0.1cm 0 0}]{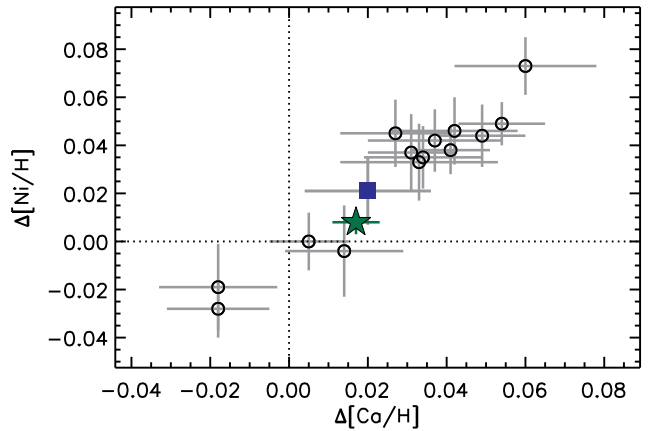}
\caption{As in Figure~\ref{f:si_fe} for the relative calcium and nickel abundances.}
\label{f:ca_ni}
\end{figure}

Admittedly, Figure~\ref{f:si_fe} corresponds to the pair of elements that have the strongest correlation (and smallest internal uncertainties), but similar results are found for most pairs of elements analyzed by L16. For example, Figure~\ref{f:ca_ni} shows the correlation between relative calcium and nickel abundances. Note also that in this case $\iota$\,Hor is an excellent fit to the Hyades open cluster trend. This is also true for all $6\leq Z\leq30$ elements.

\begin{figure}
\includegraphics[width=8.5cm, trim={0.5cm 0.6cm 0 0}]{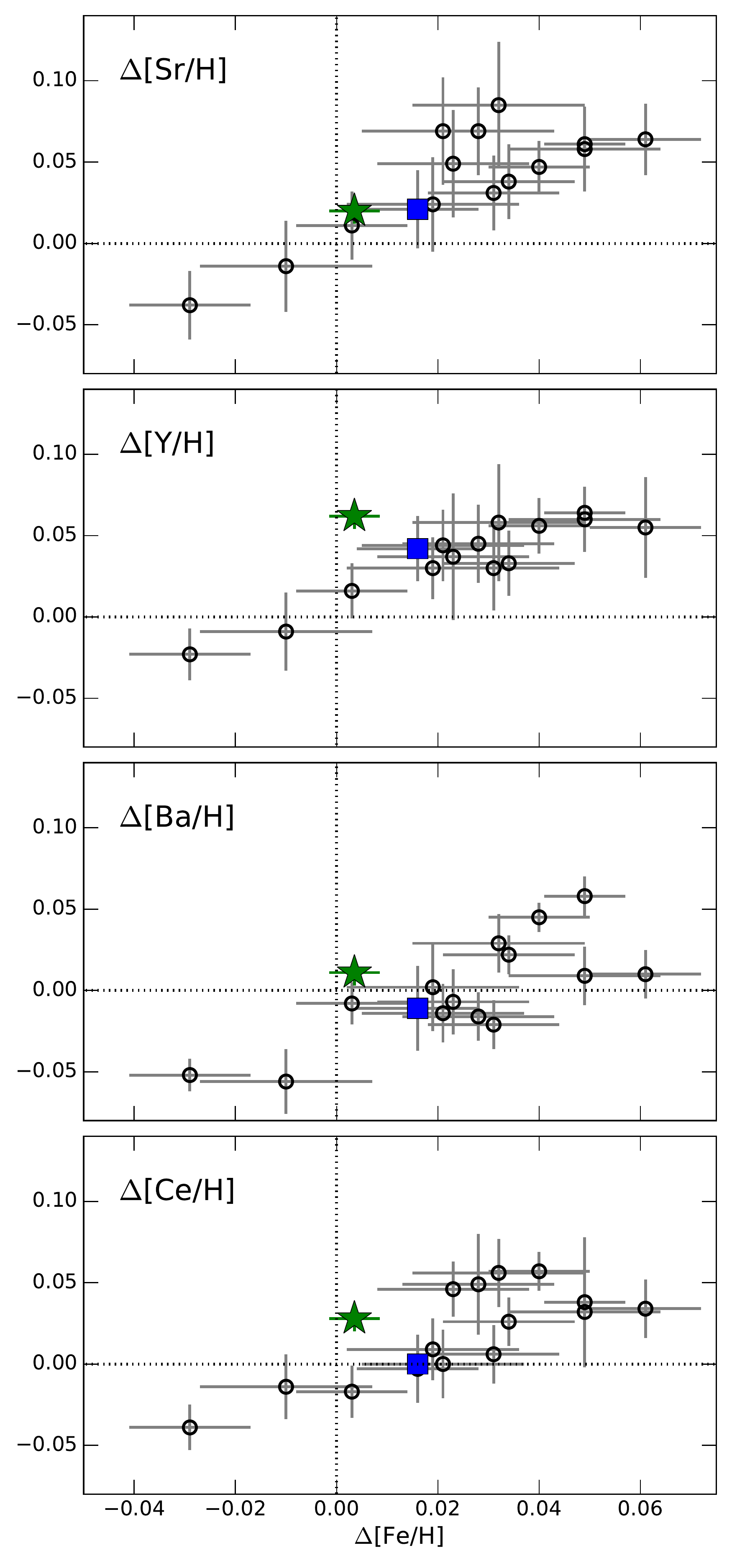}
\caption{As in Figure~\ref{f:si_fe} for relative strontium, yttrium, barium, and cerium abundances.}
\label{f:srybace}
\end{figure}

Had we only looked at the elements with atomic number below 30, excluding lithium, we would  conclude that the chemical composition of $\iota$\,Hor is indistinguishable from that of a typical Hyades star. Nevertheless, the heavier elements might cast some doubt on this statement. Figure~\ref{f:srybace} shows plots similar to those presented in Figures~\ref{f:si_fe} and \ref{f:ca_ni}, but for elements Sr, Y, Ba, and Ce. Of these four elements, only Ba was also studied by L16. For the other three, the abundances were measured in this work.

The location of $\iota$\,Hor in the four panels of Figure~\ref{f:srybace} is in excellent agreement with the Hyades trend only for Sr. Note that this element is also consistent with the low-$Z$ elements in Figure~\ref{f:delta_xh}, so this result is not at all surprising. It should also be expected that the other three elements are slightly enhanced in $\iota$\,Hor, not just relative to its Hyades twin, but in relation to the Hyades trends defined by the larger sample from L16. When $\Delta$[Fe/H] is close to zero in Figure~\ref{f:srybace}, the relative abundances of Y, Ba, and Ce in Hyades stars appear to be near zero as well, or slightly below zero. However, in the corresponding panels of Figure~\ref{f:srybace}, $\iota$\,Hor appears clearly over the zero line.

We were not able to measure Eu or Sm abundances for the L16 sample, but based on the results presented in Figure~\ref{f:srybace} and the discussion above, it would not be surprising that the corresponding plots for Eu and Sm look more like those for Y, Ba, and Ce, than the plot for Sr. After all, these two other heavy elements are also enhanced in $\iota$\,Hor relative to its Hyades twin by a similar amount ($\simeq0.02$\,dex). Thus, there are at least five species that all appear to be marginally enhanced in $\iota$\,Hor relative to the cluster.

It is not trivial to determine in general the dominant process that produces $n$-capture elements, namely the $s$- or $r$-process. Nevertheless, models like those by  \cite{bisterzo14} attempt to estimate their relative contributions for the Solar System abundances and they are often employed to categorize $n$-capture elements. An element like Ba is considered an $s$-process element because the $s$-process contribution to the Solar System Ba abundance is large, about 85\,\%, while Eu is considered an r-process element because its $s$-process contribution is only about 6\,\%. While Y, Ba, and Ce are clear $s$-process elements, Sm and Eu are not. The five elements that show slightly enhanced abundances in $\iota$\,Hor relative to its Hyades twin (Figure~\ref{f:delta_xh}) are not all $s$-process elements. Moreover, all of the heavy elements which are not enhanced in $\iota$\,Hor, namely Sr, Zr, La, and Nd have $s$-process contributions in the 60-75\,\%\ range, i.e., in between those of the Ba,Ce and Sm,Eu pairs, and about the same as Y. Thus, we find no simple explanation in terms of nucleosynthesis to justify the non-Hyades nature of $\iota$\,Hor's chemical abundance pattern. Empirically, one can conclude that unless other stars with similar marginal enhancements of these particular elements are found in the Hyades, $\iota$\,Hor does not perfectly match the chemical pattern of this cluster. While a few important studies of $s$- and $r$- process element abundances in the Hyades exist \cite[e.g.,][]{desilva07,carrera11}, we cannot employ them to tackle this problem, because these available measuments have error bars that are often greater than the full extent of the y-axis in Figure~\ref{f:delta_xh} ($\pm0.05$\,dex). The test we propose needs to be done using abundance measurements with precision errors of $\simeq0.01$\,dex, which are not yet available.

\subsection{Lithium, Rotation, and Activity} \label{s:li_prot_shk}

\begin{figure}
\includegraphics[width=8.4cm, trim={0 0.6cm 0 0}]{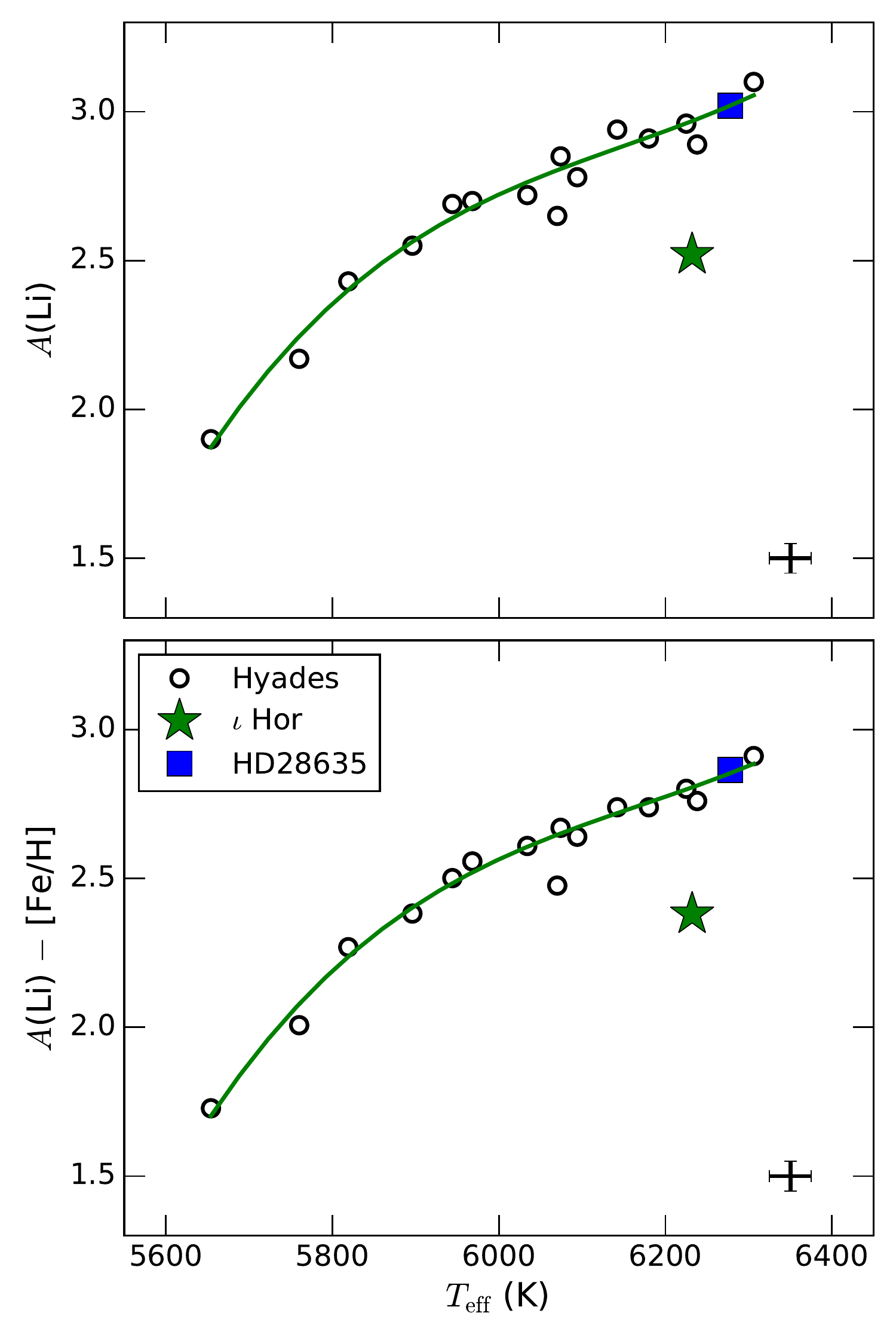}
\caption{Upper panel: lithium abundance as a function of effective temperature for Hyades stars in L16 (circles), including HD\,28635 (square). The star symbol represents $\iota$\,Hor. The solid line is a cubic fit to the Hyades data and a representative error bar is shown at bottom right. Lower panel: as in the upper panel, but for the lithium abundance minus the [Fe/H] value of each star.}
\label{f:lithium}
\end{figure}

The top panel of Figure~\ref{f:lithium} shows the lithium abundances, $A$(Li), of Hyades stars included in the L16 work as a function of the stars' effective temperatures (open circles). A very strong correlation is found between these two parameters, so we fit the data with a third-order polynomial, which is shown with a solid line in that figure. The star-to-star scatter around that fit is 0.044\,dex, after excluding the star HD\,27835, which is a marginal outlier in this plot (see below). In the bottom panel of the same Figure~\ref{f:lithium}, we show the quantity $A$(Li)--[Fe/H], that is, a lithium abundance ``corrected'' for the iron content of each star. A cubic fit is also applied to these data. Interestingly, the star-to-star scatter around this fit is significantly lower, at 0.031\,dex. This suggests that at least a small fraction of the scatter seen in $A$(Li) versus $\teff$ plots, not just of the Hyades cluster, but of late-type stars in general \cite[see, e.g., Figure~1 in][]{ramirez12:lithium}, is due to heterogeneity in the chemical composition of open clusters.

\begin{figure}
\includegraphics[width=8.7cm, trim={0.5cm 0.6cm 0 0}]{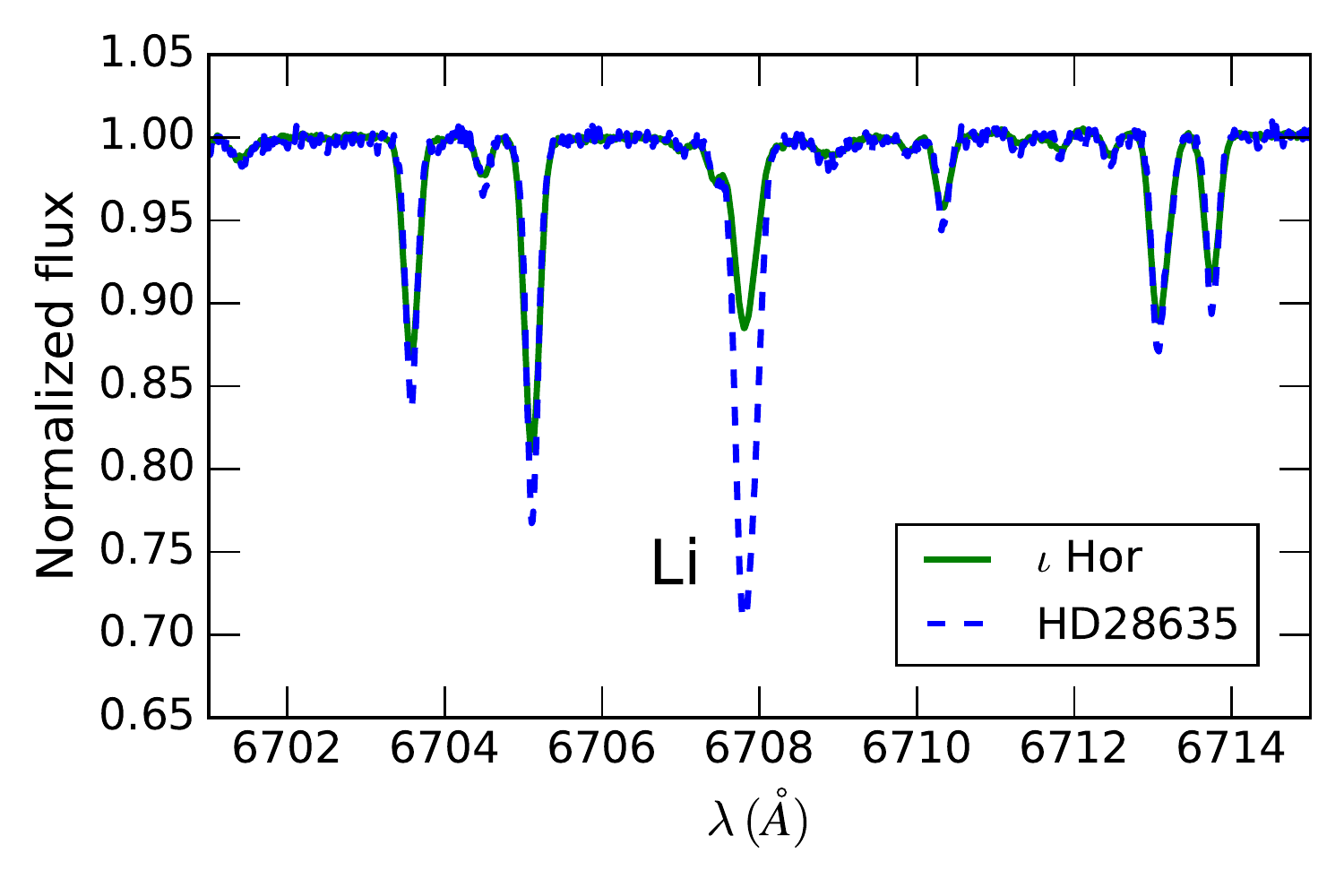}
\caption{Spectra of $\iota$\,Hor and its Hyades twin, HD\,28635, containing the 6707.8\,\AA\ lithium doublet.}
\label{f:spectra}
\end{figure}

Figure~\ref{f:lithium} shows clearly that the lithium abundance of $\iota$\,Hor is significantly lower than that of Hyades members with similar effective temperature, and therefore similar mass. $\iota$\,Hor is shown as a five-pointed star in this figure while its Hyades twin, HD\,28635, is shown with a filled square. The discrepancy is made obvious by examining the spectra themselves, as done in Figure~\ref{f:spectra}, which compares the spectra of $\iota$\,Hor and HD\,28635. This figure reveals a much weaker lithium absorption in $\iota$\,Hor relative to its twin. The small difference in effective temperature does not explain the large difference in equivalent width of the lithium doublet at 6707.8\,\AA.

Compared to the Hyades, $\iota$\,Hor has a lithium abundance that is about 0.4\,dex too low, which roughly translates to about 50\,\% of Hyades lithium content. Since the star-to-star dispersion around the $A$(Li)--$\teff$ fit is only about 0.04\,dex, this deficiency is highly significant. There is one Hyades star in L16's sample that also appears to have a low lithium abundance, HD\,27835 ($\teff=6070$\,K), but its departure from the $A$(Li)--[Fe/H] versus $\teff$ trend is only about 0.1\,dex. If $\iota$\,Hor was born in the Hyades, it must have experienced an enhanced lithium depletion.

The tight correlation between lithium abundance and $\teff$ that we have observed has been seen both in classic works \cite[e.g.,][]{boesgaard86,balachandran95} as well as in more recent studies \cite[e.g.,][]{takeda13,cummings17} of lithium abundances in the Hyades. In these works, very few, if any stars with $\teff\lesssim6300$\,K are observed to have unusually low lithium abundances. The so-called lithium dip, a region where the star-to-star lithium abundance scatter increases dramatically, begins at about 6300\,K and it could explain significantly low lithium abundances only at $\teff\simeq6500$\,K or warmer, beyond the range plotted in Figure~\ref{f:lithium}.

Lithium is depleted in cool stars because it is a fragile element that is destroyed at temperatures above $2.5\times10^6$\,K, which are already found near the base of convection zones of solar-type stars. Classical stellar evolution models fail at fully reproducing the large range of lithium abundances observed in field and cluster stars \cite[e.g.,][]{dantona94}. Non-standard mechanisms that enhance lithium depletion are thus often invoked to explain the observational data, with some success \cite[e.g.,][]{michaud86,charbonnel05,pinsonneault10}.

A number of recent studies have suggested that the presence of planets correlates with low lithium abundance \cite[e.g.,][]{israelian09,delgadomena14}, implying a connection between enhanced lithium depletion and planet formation/evolution. Although the evidence for enhanced lithium depletion in planet hosts has been argued to be a result of observational biases \cite[e.g.,][]{baumann10,ramirez12:lithium}, there are known physical mechanisms involving planets that could in principle affect the surface lithium abundance of their host stars. Since $\iota$\,Hor is known to host a gas giant planet, this is an idea worth investigating. Two mechanisms, described below, have been proposed to explain the alleged lithium-planet connection.

If a planet migrates towards its host star, the latter is expected to rotate faster to conserve angular momentum. This, in turn, can enhance mixing driven by rotation \cite[see, e.g.,][]{pinsonneault97}, which facilitates lithium depletion. In the event that a planet gets engulfed, it could replenish the star's convection zone with lithium atoms that did not burn in the planet. However, \cite{deal15} have recently argued that this ignores the important effects of ``fingering convection,'' which in fact predict additional surface lithium depletion.

Alternatively, the convection zone of a very young star that is forming planets might be locked-in by magnetic interactions with the protoplanetary disk, leading to a strong differential rotation between the stars' radiative core and its convective envelope \citep{bouvier08}. This could trigger instabilities near the base of the convection zone, enhancing the early lithium burning, particularly if the protoplanetary disk is long-lived. Contrary to the planet migration/engulfment scenario, the latter implies a slower surface rotation rate for the host star.

\begin{figure}
\includegraphics[width=8.7cm, trim={0.5cm 0.6cm 0 0}]{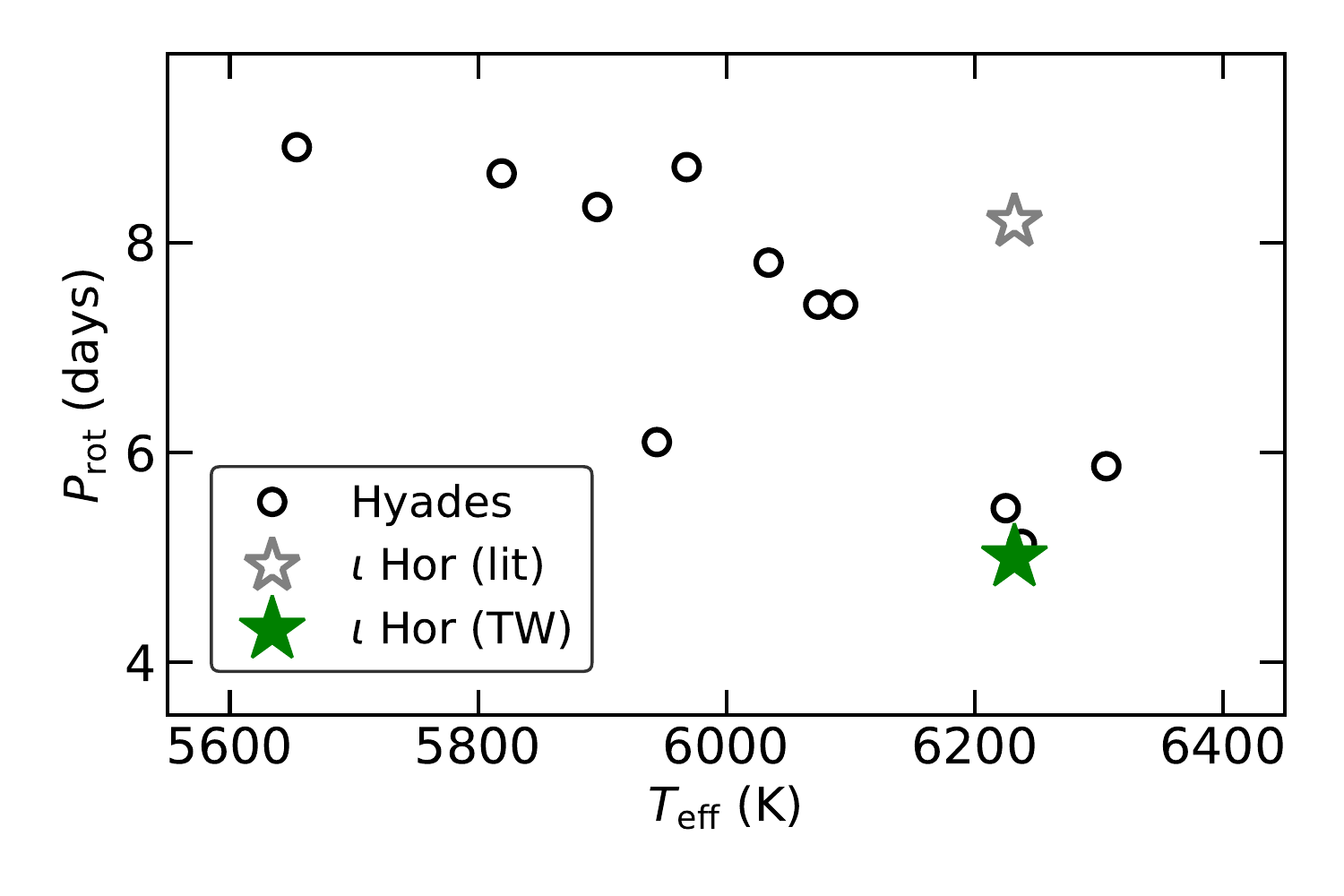}
\caption{Measured rotation periods of Hyades stars in the L16 study as a function of the stars' effective temperatures (open circles). The location of $\iota$\,Hor is shown with the star symbol. Our preferred value is shown with the green filled star symbol, while the average of previously published measurements is represented by the gray open star symbol.}
\label{f:prot}
\end{figure}

Figure~\ref{f:prot} shows the correlation between rotation period and effective temperature of the Hyades stars included in the L16 study. Not all of these stars have a measured $P_\mathrm{rot}$ value, but enough have been monitored so that an apparent trend emerges. Rotation period appears to decrease with increasing effective temperature such that it is between 8 and 9 days for stars cooler than about 6000\,K, while the three stars with $\teff\gtrsim6200$\,K all have rotation periods slightly below 6 days. The correlation is not perfect, as one star, HD\,26767, is clearly rotating too fast ($P_\mathrm{rot}=6.1$\,days) for its $\teff=5944$\,K. This star does not have a peculiar lithium abundance.

Previously-published values of the rotation period of $\iota$\,Hor are too long compared to the three Hyades stars of similar $\teff$ in Figure~\ref{f:prot}. Such slow rotation rate along with the star's unusually low lithum abundance would favor the mechanism proposed by \cite{bouvier08} to explain the enhanced lithium depletion and reject the planet migration/engulfment hypothesis. Nevertheless, as we have shown in Section~\ref{s:rotation}, the rotation period of $\iota$\,Hor is probably shorter than previously thought. In fact, the value that we derive would make $\iota$\,Hor's rotation rate consistent with that of Hyades stars of equal $\teff$. If one argues that the low lithium content of $\iota$\,Hor is somehow due to its known planet, the Hyades connection must be rejected. The fact that $\iota$\,Hor's $P_\mathrm{rot}$ matches the Hyades trend might simply indicate that they have a similar age, as suggested by \cite{vauclair08}.

We should emphasize that the result discussed above is basically determined by three Hyades stars with $\teff\gtrsim6\,200$\,K and $P_\mathrm{rot}\lesssim6$\,days. Although the zero point of the effective temperatures used in this work, which are from L16, is uncertain, the {\it relative} $\teff$ values have precision errors of only about 10\,K. Thus, these three stars are in fact very similar in $\teff$ to $\iota$\,Hor. Their rotation periods have been measured by one author only, namely A.\ Kundert, with the exception of HD\,27406 ($P_\mathrm{rot}=5.47$\,days), which has a previous $P_\mathrm{rot}=5.45\pm0.02$\,days measurement from \cite{radick87} that is in excellent agreement with that derived by Kundert. Thus, the data employed are very reliable, suggesting that this is a robust observational result.

If our estimate of the rotation period of $\iota$\,Hor were incorrect and the literature value were more precise, one could note that a low lithium abundance and a slower rotation period are a natural consequence of standard stellar evolution. Older stars have had more time to lose angular momentum through winds and deplete more lithium than their younger counterparts. Thus, in that case one could argue that $\iota$\,Hor is simply older than the Hyades.


For stars which have otherwise identical fundamental properties, \cite{skumanich72} predicts a $P_\mathrm{rot}\propto\tau^{0.5}$ gyrochronology law, where $\tau$ is the star's age (see also \citealt{barnes07,barnes10}). If $\iota$\,Hor were a ``normal'' Hyades star in its rotation, then its $P_\mathrm{rot}$ would be about 5\,days. If we rely on previously published works that suggest a $P_\mathrm{rot}$ value of about 8.2\,days, we would infer a gyrochronological age of $\tau=\tau(\mathrm{Hyades})\times(8.2/5.0)^{1/0.5}=1.7$\,Gyr (assuming an age of 650\,Myr for the Hyades). 

\newpage

\subsection{Dynamical Analysis}

\begin{figure}
\centering
\includegraphics[width=7.5cm, trim={0.5cm 0.4cm 0 0}]{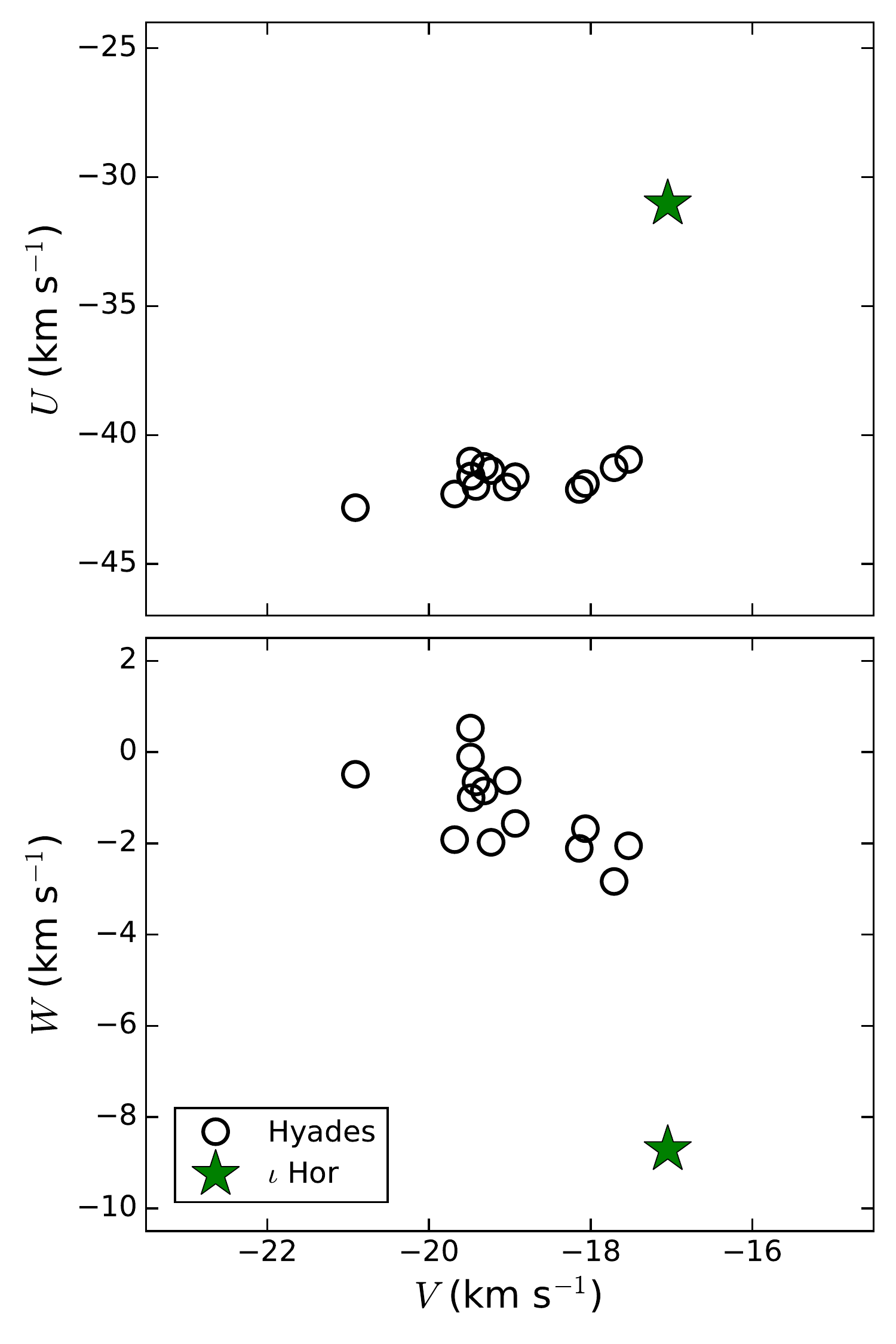}
\caption{Galactic space velocities of the L16's Hyades stars (circles) and $\iota$\,Hor (star symbol).}
\label{f:uvw}
\end{figure}

The Galactic space velocities $U,V,W$ (heliocentric) of $\iota$\,Hor are close, but not within the distribution of velocities of Hyades stars, in particular the stars of the L16 study,\footnote{One of the stars in L16 (HD\,27835) does not have a {\it Hipparcos} parallax available, therefore it is not included in this dynamical analysis.} as shown in Figure~\ref{f:uvw}. These velocities were computed in the same manner for all stars using the \texttt{galpy} library for Galactic dynamics \citep{bovy15}.\footnote{\url{http://github.com/jobovy/galpy}} Since all these stars have precise trigonometric parallaxes, proper motions, and radial velocities, the $U,V,W$ values have errors of order 0.5\,km\,s$^{-1}$.

If superimposed on the countour diagrams for the mean covariance matrix of space velocities for the large sample of Hyades stars by \citet[][their Figure~16]{perryman98}, $\iota$\,Hor would fall outside of the 99.99\,\% confidence level countour. Thus, strictly speaking, $\iota$\,Hor does not presently move in the same direction as the Hyades cluster. In order to further examine whether these kinematic properties support the idea of a common origin with the Hyades, we computed the stars' orbits in a Galactic potential using \texttt{galpy} and followed the logic from \cite{bobylev11} and \cite{ramirez14:siblings} for finding stellar siblings.

\begin{figure}
\centering
\includegraphics[width=7.5cm, trim={0.5cm 0.6cm 0 0}]{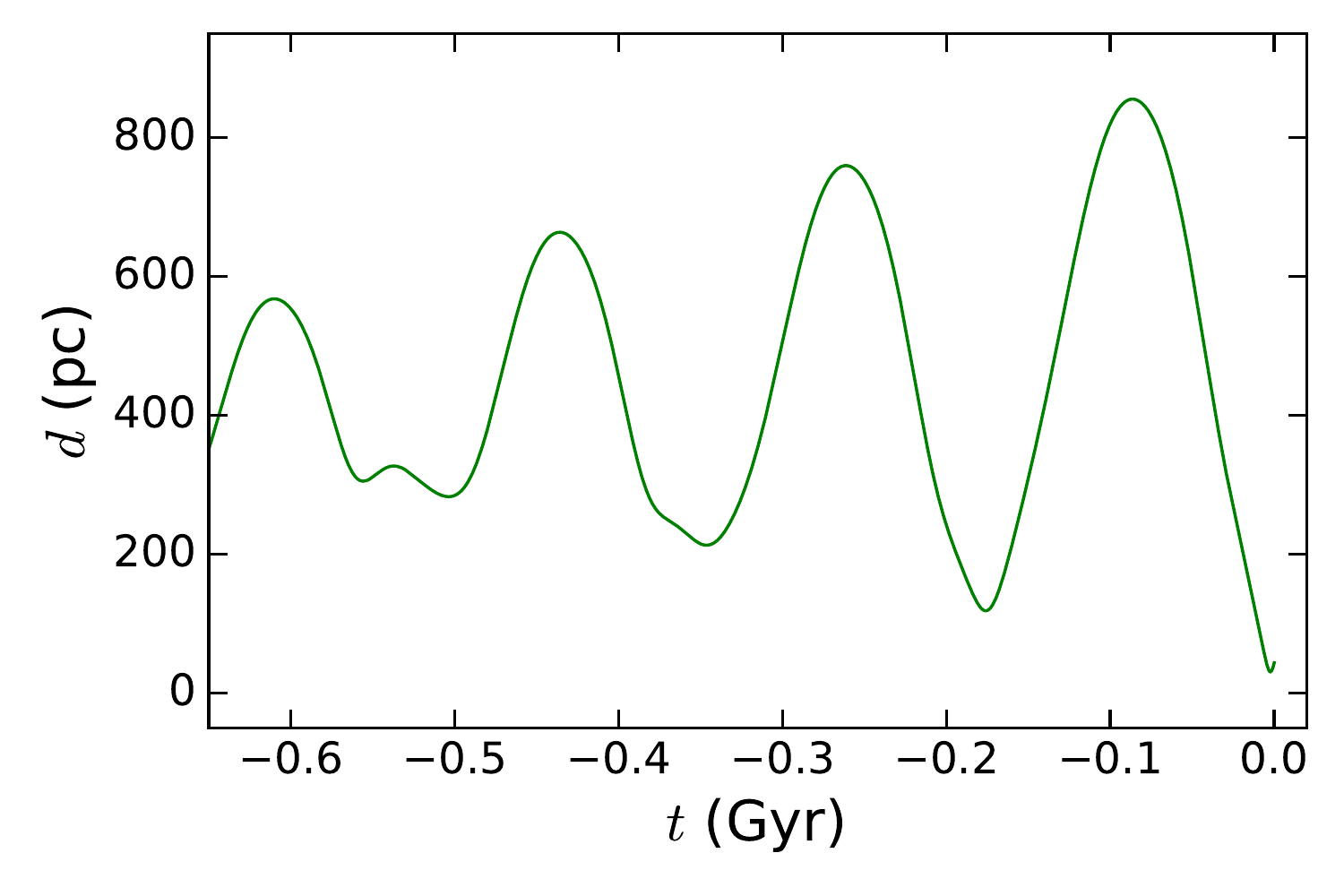}
\caption{Distance between $\iota$\,Hor and the average location of L16's Hyades stars as a funcion of time in the past according to our Galactic orbit calculations.}
\label{f:dvt}
\end{figure}

We computed the orbits of $\iota$\,Hor and 15 of the Hyades stars in L16 backwards in time for 650\,Myr using the ``simple, easy-to-use'' Milky Way potential provided in \texttt{galpy} (MWPotential2014). This model consists of a halo, bulge, and disk, but it does not include spiral arms or a central supermassive black hole. The details of the model and the parameters employed are listed in Table~1 of \citeauthor{bovy15}'s paper.  At every time step in the simulation, we calculated the distance between $\iota$\,Hor and the average location of L16's Hyades stars. The result, which is shown in Figure~\ref{f:dvt}, suggests that even though today $\iota$\,Hor moves in the same general direction as the Hyades supercluster, in the past, it may have been farther away from the cluster itself. In fact, the farther back in time this simulation is run, the more $\iota$\,Hor separates from the mean location of the Hyades. Figure~\ref{f:dvt} implies that at the time the Hyades formed, 650\,Myr ago, $\iota$\,Hor was nearly 400\,pc away from the cluster, likely too far to have formed inside the cluster, but arguably within the same spiral arm.

Admittedly, computations of Galactic orbits for individual stars are highly uncertain. In particular, significant differences in the orbital properties would be obtained if the potential from spiral arms were included \citep{mishurov11,martinez-barbosa16}. Nevertheless, one would expect our relatively simple calculation to provide the most optimistic result in this context.

\section{Conclusion}

One could say that the detailed chemical composition of $\iota$\,Hor is consistent with that of the Hyades if only elements with atomic number below 30, excluding lithium, are investigated. This conclusion can be promptly made when employing abundance measurements of ordinary accuracy, but we find it also valid in our high-precision work. Nevertheless, we also find that the abundances of about half of all heavy elements ($Z>30$) analyzed are slightly enhanced in $\iota$\,Hor relative to the Hyades. The latter can only be confirmed when elemental abundances are measured with a precision of 0.02\,dex or better.

The lithium abundance of $\iota$\,Hor is too low compared to Hyades stars of similar effective temperature. If a true member of this cluster, $\iota$\,Hor's enhanced lithium depletion could be attributed to its planet, but this would require the star to have a peculiar (non-Hyades-like) rotation rate, which is not observed beyond doubt. Furthermore, Galactic orbit calculations suggest that $\iota$\,Hor was far away from the cluster when it formed.

The suggestion of a $\iota$\,Hor--Hyades connection by \cite{vauclair08} was motivated by the star's kinematic properties, even though only one of the two classic criteria of spatial convergence with the cluster is actually satisfied by this star. The proposed association was argued to be confirmed once a helium abundance measurement was made and reported to be in good agreement with that of the Hyades. However, we note that the helium content of the Universe has not changed significantly since the Big Bang and it therefore provides only a very weak constraint for the common origin hypothesis. Indeed, in the time that metallicity has increased by a factor of 10 (1.00\,dex), the helium abundance has increased by about 15\,\%\ \cite[0.06\,dex; see, e.g., Figure 3 in][]{balser06}. Moreover, since the helium abundance of $\iota$\,Hor was measured very precisely by \citeauthor{vauclair08} using acoustic oscillation data, but that of the Hyades was estimated using a different method, namely model fits to mass-luminosity relations of binaries, the agreement might well be purely coincidental.

If not a Hyades member, which appears to be the favored scenario given the data that we have analyzed, our results constitute a challenge for present-day chemical tagging efforts. The latter are capable of measuring abundances with an optimistic uncertainty of 0.05\,dex. With those errors, $\iota$\,Hor would be easily associated, chemically, with the Hyades. However, upon careful analysis, it seems quite unlikely that $\iota$\,Hor was born in the Hyades cluster.

\acknowledgements

DY acknowledges support from the Australian Research Council (FT140100554). A portion of this work was carried out by EG during the TAURUS program at The University of Texas at Austin, which was partly supported by the National Science Foundation under grant 1313075.  DLL thanks the Robert A. Welch foundation of Houston, Texas for support through grant F-634. Some of this work is based on observations made with ESO Telescopes at the La Silla Paranal Observatory under the programs listed in Table~\ref{t:archive} and Section~\ref{s:data}.


\newpage

\begin{thebibliography}{}
\expandafter\ifx\csname natexlab\endcsname\relax\def\natexlab#1{#1}\fi

\bibitem[{{Adibekyan} {et~al.}(2012){Adibekyan}, {Sousa}, {Santos}, {Delgado
  Mena}, {Gonz{\'a}lez Hern{\'a}ndez}, {Israelian}, {Mayor}, \&
  {Khachatryan}}]{adibekyan12}
{Adibekyan}, V.~Z., {Sousa}, S.~G., {Santos}, N.~C., {et~al.} 2012, \aap, 545,
  A32

\bibitem[{{Asplund}(2005)}]{asplund05:review}
{Asplund}, M. 2005, \araa, 43, 481

\bibitem[{{Balachandran}(1995)}]{balachandran95}
{Balachandran}, S. 1995, \apj, 446, 203

\bibitem[{{Balser}(2006)}]{balser06}
{Balser}, D.~S. 2006, \aj, 132, 2326

\bibitem[{{Barnes}(2007)}]{barnes07}
{Barnes}, S.~A. 2007, \apj, 669, 1167

\bibitem[{{Barnes}(2010)}]{barnes10}
---. 2010, \apj, 722, 222

\bibitem[{{Baumann} {et~al.}(2010){Baumann}, {Ram{\'{\i}}rez}, {Mel{\'e}ndez},
  {Asplund}, \& {Lind}}]{baumann10}
{Baumann}, P., {Ram{\'{\i}}rez}, I., {Mel{\'e}ndez}, J., {Asplund}, M., \&
  {Lind}, K. 2010, \aap, 519, A87

\bibitem[{{Bedell} {et~al.}(2014){Bedell}, {Mel{\'e}ndez}, {Bean},
  {Ram{\'{\i}}rez}, {Leite}, \& {Asplund}}]{bedell14}
{Bedell}, M., {Mel{\'e}ndez}, J., {Bean}, J.~L., {et~al.} 2014, \apj, 795, 23

\bibitem[{{Bensby} {et~al.}(2014){Bensby}, {Feltzing}, \& {Oey}}]{bensby14}
{Bensby}, T., {Feltzing}, S., \& {Oey}, M.~S. 2014, \aap, 562, A71

\bibitem[{{Bisterzo} {et~al.}(2014){Bisterzo}, {Travaglio}, {Gallino},
  {Wiescher}, \& {K{\"a}ppeler}}]{bisterzo14}
{Bisterzo}, S., {Travaglio}, C., {Gallino}, R., {Wiescher}, M., \&
  {K{\"a}ppeler}, F. 2014, \apj, 787, 10

\bibitem[{{Bobylev} {et~al.}(2011){Bobylev}, {Bajkova}, {Myll{\"a}ri}, \&
  {Valtonen}}]{bobylev11}
{Bobylev}, V.~V., {Bajkova}, A.~T., {Myll{\"a}ri}, A., \& {Valtonen}, M. 2011,
  Astronomy Letters, 37, 550

\bibitem[{{Boesgaard} \& {Tripicco}(1986)}]{boesgaard86}
{Boesgaard}, A.~M., \& {Tripicco}, M.~J. 1986, \apjl, 302, L49

\bibitem[{{Boisse} {et~al.}(2011){Boisse}, {Bouchy}, {H{\'e}brard}, {Bonfils},
  {Santos}, \& {Vauclair}}]{boisse11}
{Boisse}, I., {Bouchy}, F., {H{\'e}brard}, G., {et~al.} 2011, \aap, 528, A4

\bibitem[{{Bouvier}(2008)}]{bouvier08}
{Bouvier}, J. 2008, \aap, 489, L53

\bibitem[{{Bovy}(2015)}]{bovy15}
{Bovy}, J. 2015, \apjs, 216, 29

\bibitem[{{Brucalassi} {et~al.}(2014){Brucalassi}, {Pasquini}, {Saglia},
  {Ruiz}, {Bonifacio}, {Bedin}, {Biazzo}, {Melo}, {Lovis}, \&
  {Randich}}]{brucalassi14}
{Brucalassi}, A., {Pasquini}, L., {Saglia}, R., {et~al.} 2014, \aap, 561, L9

\bibitem[{{Carrera} \& {Pancino}(2011)}]{carrera11}
{Carrera}, R., \& {Pancino}, E. 2011, \aap, 535, A30

\bibitem[{{Castelli} \& {Kurucz}(2003)}]{castelli03}
{Castelli}, F., \& {Kurucz}, R.~L. 2003, in IAU Symposium, Vol. 210, Modelling
  of Stellar Atmospheres, ed. N.~{Piskunov}, W.~W. {Weiss}, \& D.~F. {Gray}, 20

\bibitem[{{Charbonnel} \& {Talon}(2005)}]{charbonnel05}
{Charbonnel}, C., \& {Talon}, S. 2005, Science, 309, 2189

\bibitem[{{Clarkson} {et~al.}(2003){Clarkson}, {Charles}, {Coe}, {Laycock},
  {Tout}, \& {Wilson}}]{clarkson03}
{Clarkson}, W.~I., {Charles}, P.~A., {Coe}, M.~J., {et~al.} 2003, \mnras, 339,
  447

\bibitem[{{Cummings} {et~al.}(2017){Cummings}, {Deliyannis}, {Maderak}, \&
  {Steinhauer}}]{cummings17}
{Cummings}, J.~D., {Deliyannis}, C.~P., {Maderak}, R.~M., \& {Steinhauer}, A.
  2017, \aj, 153, 128

\bibitem[{{D'Antona} \& {Mazzitelli}(1994)}]{dantona94}
{D'Antona}, F., \& {Mazzitelli}, I. 1994, \apjs, 90, 467

\bibitem[{{De Gennaro} {et~al.}(2009){De Gennaro}, {von Hippel}, {Jefferys},
  {Stein}, {van Dyk}, \& {Jeffery}}]{degennaro09}
{De Gennaro}, S., {von Hippel}, T., {Jefferys}, W.~H., {et~al.} 2009, \apj,
  696, 12

\bibitem[{{De Silva} {et~al.}(2007){De Silva}, {Freeman}, {Bland-Hawthorn},
  {Asplund}, \& {Bessell}}]{desilva07}
{De Silva}, G.~M., {Freeman}, K.~C., {Bland-Hawthorn}, J., {Asplund}, M., \&
  {Bessell}, M.~S. 2007, \aj, 133, 694

\bibitem[{{Deal} {et~al.}(2015){Deal}, {Richard}, \& {Vauclair}}]{deal15}
{Deal}, M., {Richard}, O., \& {Vauclair}, S. 2015, \aap, 584, A105

\bibitem[{{Delgado Mena} {et~al.}(2014){Delgado Mena}, {Israelian},
  {Gonz{\'a}lez Hern{\'a}ndez}, {Sousa}, {Mortier}, {Santos}, {Adibekyan},
  {Fernandes}, {Rebolo}, {Udry}, \& {Mayor}}]{delgadomena14}
{Delgado Mena}, E., {Israelian}, G., {Gonz{\'a}lez Hern{\'a}ndez}, J.~I.,
  {et~al.} 2014, \aap, 562, A92

\bibitem[{{Eggen}(1995)}]{eggen95}
{Eggen}, O.~J. 1995, \aj, 110, 2862

\bibitem[{{Famaey} {et~al.}(2007){Famaey}, {Pont}, {Luri}, {Udry}, {Mayor}, \&
  {Jorissen}}]{famaey07}
{Famaey}, B., {Pont}, F., {Luri}, X., {et~al.} 2007, \aap, 461, 957

\bibitem[{{Flores} {et~al.}(2017){Flores}, {Buccino}, {Saffe}, \&
  {Mauas}}]{flores17}
{Flores}, M.~G., {Buccino}, A.~P., {Saffe}, C.~E., \& {Mauas}, P.~J.~D. 2017,
  \mnras, 464, 4299

\bibitem[{{Freeman} \& {Bland-Hawthorn}(2002)}]{freeman02}
{Freeman}, K., \& {Bland-Hawthorn}, J. 2002, \araa, 40, 487

\bibitem[{{Gonz{\'a}lez Hern{\'a}ndez} {et~al.}(2013){Gonz{\'a}lez
  Hern{\'a}ndez}, {Delgado-Mena}, {Sousa}, {Israelian}, {Santos}, {Adibekyan},
  \& {Udry}}]{gonzalez-hernandez13}
{Gonz{\'a}lez Hern{\'a}ndez}, J.~I., {Delgado-Mena}, E., {Sousa}, S.~G.,
  {et~al.} 2013, \aap, 552, A6

\bibitem[{{Hathaway}(2015)}]{hathaway15}
{Hathaway}, D.~H. 2015, Living Reviews in Solar Physics, 12, 4

\bibitem[{{Hogg} {et~al.}(2016){Hogg}, {Casey}, {Ness}, {Rix},
  {Foreman-Mackey}, {Hasselquist}, {Ho}, {Holtzman}, {Majewski}, {Martell},
  {M{\'e}sz{\'a}ros}, {Nidever}, \& {Shetrone}}]{hogg16}
{Hogg}, D.~W., {Casey}, A.~R., {Ness}, M., {et~al.} 2016, \apj, 833, 262

\bibitem[{{Horne} \& {Baliunas}(1986)}]{horne86}
{Horne}, J.~H., \& {Baliunas}, S.~L. 1986, \apj, 302, 757

\bibitem[{{Israelian} {et~al.}(2009){Israelian}, {Delgado Mena}, {Santos},
  {Sousa}, {Mayor}, {Udry}, {Dom{\'{\i}}nguez Cerde{\~n}a}, {Rebolo}, \&
  {Randich}}]{israelian09}
{Israelian}, G., {Delgado Mena}, E., {Santos}, N.~C., {et~al.} 2009, \nat, 462,
  189

\bibitem[{{Janes} {et~al.}(1988){Janes}, {Tilley}, \& {Lynga}}]{janes88}
{Janes}, K.~A., {Tilley}, C., \& {Lynga}, G. 1988, \aj, 95, 771

\bibitem[{{K{\"u}rster} {et~al.}(2000){K{\"u}rster}, {Endl}, {Els}, {Hatzes},
  {Cochran}, {D{\"o}bereiner}, \& {Dennerl}}]{kurster00}
{K{\"u}rster}, M., {Endl}, M., {Els}, S., {et~al.} 2000, \aap, 353, L33

\bibitem[{{Lebreton} {et~al.}(2001){Lebreton}, {Fernandes}, \&
  {Lejeune}}]{lebreton01}
{Lebreton}, Y., {Fernandes}, J., \& {Lejeune}, T. 2001, \aap, 374, 540

\bibitem[{{Lind} {et~al.}(2009){Lind}, {Asplund}, \& {Barklem}}]{lind09}
{Lind}, K., {Asplund}, M., \& {Barklem}, P.~S. 2009, \aap, 503, 541

\bibitem[{{Liu} {et~al.}(2016){Liu}, {Yong}, {Asplund}, {Ram{\'{\i}}rez}, \&
  {Mel{\'e}ndez}}]{liu16}
{Liu}, F., {Yong}, D., {Asplund}, M., {Ram{\'{\i}}rez}, I., \& {Mel{\'e}ndez},
  J. 2016, \mnras, 457, 3934

\bibitem[{{Lomb}(1976)}]{lomb76}
{Lomb}, N.~R. 1976, \apss, 39, 447

\bibitem[{{Malavolta} {et~al.}(2016){Malavolta}, {Nascimbeni}, {Piotto},
  {Quinn}, {Borsato}, {Granata}, {Bonomo}, {Marzari}, {Bedin}, {Rainer},
  {Desidera}, {Lanza}, {Poretti}, {Sozzetti}, {White}, {Latham}, {Cunial},
  {Libralato}, {Nardiello}, {Boccato}, {Claudi}, {Cosentino}, {Covino},
  {Gratton}, {Maggio}, {Micela}, {Molinari}, {Pagano}, {Smareglia}, {Affer},
  {Andreuzzi}, {Aparicio}, {Benatti}, {Bignamini}, {Borsa}, {Damasso}, {Di
  Fabrizio}, {Harutyunyan}, {Esposito}, {Fiorenzano}, {Gandolfi}, {Giacobbe},
  {Gonz{\'a}lez Hern{\'a}ndez}, {Maldonado}, {Masiero}, {Molinaro}, {Pedani},
  \& {Scandariato}}]{malavolta16}
{Malavolta}, L., {Nascimbeni}, V., {Piotto}, G., {et~al.} 2016, \aap, 588, A118

\bibitem[{{Mann} {et~al.}(2016){Mann}, {Gaidos}, {Mace}, {Johnson}, {Bowler},
  {LaCourse}, {Jacobs}, {Vanderburg}, {Kraus}, {Kaplan}, \& {Jaffe}}]{mann16}
{Mann}, A.~W., {Gaidos}, E., {Mace}, G.~N., {et~al.} 2016, \apj, 818, 46

\bibitem[{{Mart{\'{\i}}nez-Barbosa} {et~al.}(2016){Mart{\'{\i}}nez-Barbosa},
  {Brown}, {Boekholt}, {Portegies Zwart}, {Antiche}, \&
  {Antoja}}]{martinez-barbosa16}
{Mart{\'{\i}}nez-Barbosa}, C.~A., {Brown}, A.~G.~A., {Boekholt}, T., {et~al.}
  2016, \mnras, 457, 1062

\bibitem[{{Mel{\'e}ndez} {et~al.}(2009){Mel{\'e}ndez}, {Asplund}, {Gustafsson},
  \& {Yong}}]{melendez09:twins}
{Mel{\'e}ndez}, J., {Asplund}, M., {Gustafsson}, B., \& {Yong}, D. 2009, \apjl,
  704, L66

\bibitem[{{Mel{\'e}ndez} {et~al.}(2014){Mel{\'e}ndez}, {Ram{\'{\i}}rez},
  {Karakas}, {Yong}, {Monroe}, {Bedell}, {Bergemann}, {Asplund}, {Tucci Maia},
  {Bean}, {do Nascimento}, {Bazot}, {Alves-Brito}, {Freitas}, \&
  {Castro}}]{melendez14:18sco}
{Mel{\'e}ndez}, J., {Ram{\'{\i}}rez}, I., {Karakas}, A.~I., {et~al.} 2014,
  \apj, 791, 14

\bibitem[{{Metcalfe} {et~al.}(2010){Metcalfe}, {Basu}, {Henry}, {Soderblom},
  {Judge}, {Kn{\"o}lker}, {Mathur}, \& {Rempel}}]{metcalfe10}
{Metcalfe}, T.~S., {Basu}, S., {Henry}, T.~J., {et~al.} 2010, \apjl, 723, L213

\bibitem[{{Michaud}(1986)}]{michaud86}
{Michaud}, G. 1986, \apj, 302, 650

\bibitem[{{Mishurov} \& {Acharova}(2011)}]{mishurov11}
{Mishurov}, Y.~N., \& {Acharova}, I.~A. 2011, \mnras, 412, 1771

\bibitem[{{Montes} {et~al.}(2001){Montes}, {L{\'o}pez-Santiago}, {G{\'a}lvez},
  {Fern{\'a}ndez-Figueroa}, {De Castro}, \& {Cornide}}]{montes01}
{Montes}, D., {L{\'o}pez-Santiago}, J., {G{\'a}lvez}, M.~C., {et~al.} 2001,
  \mnras, 328, 45

\bibitem[{{Nissen}(2015)}]{nissen15}
{Nissen}, P.~E. 2015, \aap, 579, A52

\bibitem[{{Paulson} {et~al.}(2004){Paulson}, {Cochran}, \&
  {Hatzes}}]{paulson04}
{Paulson}, D.~B., {Cochran}, W.~D., \& {Hatzes}, A.~P. 2004, \aj, 127, 3579

\bibitem[{{Perryman} {et~al.}(1998){Perryman}, {Brown}, {Lebreton}, {Gomez},
  {Turon}, {Cayrel de Strobel}, {Mermilliod}, {Robichon}, {Kovalevsky}, \&
  {Crifo}}]{perryman98}
{Perryman}, M.~A.~C., {Brown}, A.~G.~A., {Lebreton}, Y., {et~al.} 1998, \aap,
  331, 81

\bibitem[{{Pinsonneault}(1997)}]{pinsonneault97}
{Pinsonneault}, M. 1997, \araa, 35, 557

\bibitem[{{Pinsonneault}(2010)}]{pinsonneault10}
{Pinsonneault}, M.~H. 2010, in IAU Symposium, Vol. 268, IAU Symposium, ed.
  {C.~Charbonnel, M.~Tosi, F.~Primas, \& C.~Chiappini}, 375--380

\bibitem[{{Pomp{\'e}ia} {et~al.}(2011){Pomp{\'e}ia}, {Masseron}, {Famaey}, {van
  Eck}, {Jorissen}, {Minchev}, {Siebert}, {Sneden}, {L{\'e}pine}, {Siopis},
  {Gentile}, {Dermine}, {Pasquato}, {van Winckel}, {Waelkens}, {Raskin},
  {Prins}, {Pessemier}, {Hensberge}, {Fr{\'e}mat}, {Dumortier}, \&
  {Bienaym{\'e}}}]{pompeia11}
{Pomp{\'e}ia}, L., {Masseron}, T., {Famaey}, B., {et~al.} 2011, \mnras, 415,
  1138

\bibitem[{{Quinn} {et~al.}(2012){Quinn}, {White}, {Latham}, {Buchhave},
  {Cantrell}, {Dahm}, {F{\H u}r{\'e}sz}, {Szentgyorgyi}, {Geary}, {Torres},
  {Bieryla}, {Berlind}, {Calkins}, {Esquerdo}, \& {Stefanik}}]{quinn12}
{Quinn}, S.~N., {White}, R.~J., {Latham}, D.~W., {et~al.} 2012, \apjl, 756, L33

\bibitem[{{Quinn} {et~al.}(2014){Quinn}, {White}, {Latham}, {Buchhave},
  {Torres}, {Stefanik}, {Berlind}, {Bieryla}, {Calkins}, {Esquerdo}, {F{\H
  u}r{\'e}sz}, {Geary}, \& {Szentgyorgyi}}]{quinn14}
---. 2014, \apj, 787, 27

\bibitem[{{Radick} {et~al.}(1987){Radick}, {Thompson}, {Lockwood}, {Duncan}, \&
  {Baggett}}]{radick87}
{Radick}, R.~R., {Thompson}, D.~T., {Lockwood}, G.~W., {Duncan}, D.~K., \&
  {Baggett}, W.~E. 1987, \apj, 321, 459

\bibitem[{{Ram{\'{\i}}rez} {et~al.}(2012){Ram{\'{\i}}rez}, {Fish}, {Lambert},
  \& {Allende Prieto}}]{ramirez12:lithium}
{Ram{\'{\i}}rez}, I., {Fish}, J.~R., {Lambert}, D.~L., \& {Allende Prieto}, C.
  2012, \apj, 756, 46

\bibitem[{{Ram{\'{\i}}rez} {et~al.}(2014{\natexlab{a}}){Ram{\'{\i}}rez},
  {Mel{\'e}ndez}, \& {Asplund}}]{ramirez14:bst}
{Ram{\'{\i}}rez}, I., {Mel{\'e}ndez}, J., \& {Asplund}, M. 2014{\natexlab{a}},
  \aap, 561, A7

\bibitem[{{Ram{\'{\i}}rez} {et~al.}(2014{\natexlab{b}}){Ram{\'{\i}}rez},
  {Bajkova}, {Bobylev}, {Roederer}, {Lambert}, {Endl}, {Cochran}, {MacQueen},
  \& {Wittenmyer}}]{ramirez14:siblings}
{Ram{\'{\i}}rez}, I., {Bajkova}, A.~T., {Bobylev}, V.~V., {et~al.}
  2014{\natexlab{b}}, \apj, 787, 154

\bibitem[{{Ram{\'{\i}}rez} {et~al.}(2014{\natexlab{c}}){Ram{\'{\i}}rez},
  {Mel{\'e}ndez}, {Bean}, {Asplund}, {Bedell}, {Monroe}, {Casagrande},
  {Schirbel}, {Dreizler}, {Teske}, {Tucci Maia}, {Alves-Brito}, \&
  {Baumann}}]{ramirez14:harps}
{Ram{\'{\i}}rez}, I., {Mel{\'e}ndez}, J., {Bean}, J., {et~al.}
  2014{\natexlab{c}}, \aap, 572, A48

\bibitem[{{Ram{\'{\i}}rez} {et~al.}(2015){Ram{\'{\i}}rez}, {Khanal}, {Aleo},
  {Sobotka}, {Liu}, {Casagrande}, {Mel{\'e}ndez}, {Yong}, {Lambert}, \&
  {Asplund}}]{ramirez15}
{Ram{\'{\i}}rez}, I., {Khanal}, S., {Aleo}, P., {et~al.} 2015, \apj, 808, 13

\bibitem[{{Reddy} {et~al.}(2006){Reddy}, {Lambert}, \& {Allende
  Prieto}}]{reddy06}
{Reddy}, B.~E., {Lambert}, D.~L., \& {Allende Prieto}, C. 2006, \mnras, 367,
  1329

\bibitem[{{Saar} {et~al.}(1997){Saar}, {Huovelin}, {Osten}, \&
  {Shcherbakov}}]{saar97:helium}
{Saar}, S.~H., {Huovelin}, J., {Osten}, R.~A., \& {Shcherbakov}, A.~G. 1997,
  \aap, 326, 741

\bibitem[{{Saar} \& {Osten}(1997)}]{saar97:south}
{Saar}, S.~H., \& {Osten}, R.~A. 1997, \mnras, 284, 803

\bibitem[{{Sanz-Forcada} {et~al.}(2013){Sanz-Forcada}, {Stelzer}, \&
  {Metcalfe}}]{sanz-forcada13}
{Sanz-Forcada}, J., {Stelzer}, B., \& {Metcalfe}, T.~S. 2013, \aap, 553, L6

\bibitem[{{Scargle}(1982)}]{scargle82}
{Scargle}, J.~D. 1982, \apj, 263, 835

\bibitem[{{Skumanich}(1972)}]{skumanich72}
{Skumanich}, A. 1972, \apj, 171, 565

\bibitem[{{Sneden}(1973)}]{sneden73}
{Sneden}, C.~A. 1973, PhD thesis, The University of Texas at Austin

\bibitem[{{Spina} {et~al.}(2016){Spina}, {Mel{\'e}ndez}, \&
  {Ram{\'{\i}}rez}}]{spina16}
{Spina}, L., {Mel{\'e}ndez}, J., \& {Ram{\'{\i}}rez}, I. 2016, \aap, 585, A152

\bibitem[{{Takeda} {et~al.}(2013){Takeda}, {Honda}, {Ohnishi}, {Ohkubo},
  {Hirata}, \& {Sadakane}}]{takeda13}
{Takeda}, Y., {Honda}, S., {Ohnishi}, T., {et~al.} 2013, \pasj, 65, 53

\bibitem[{{Vauclair} {et~al.}(2008){Vauclair}, {Laymand}, {Bouchy}, {Vauclair},
  {Hui Bon Hoa}, {Charpinet}, \& {Bazot}}]{vauclair08}
{Vauclair}, S., {Laymand}, M., {Bouchy}, F., {et~al.} 2008, \aap, 482, L5

\bibitem[{{Zechmeister} {et~al.}(2013){Zechmeister}, {K{\"u}rster}, {Endl}, {Lo
  Curto}, {Hartman}, {Nilsson}, {Henning}, {Hatzes}, \&
  {Cochran}}]{zechmeister13}
{Zechmeister}, M., {K{\"u}rster}, M., {Endl}, M., {et~al.} 2013, \aap, 552, A78

\end{thebibliography}
\end{document}